\documentclass[notitlepage,superscriptaddress,reprint,amsmath,amssymb,aps,prb]{revtex4-2}

\usepackage{graphicx,hyperref,xcolor} 
\hypersetup{colorlinks=true, linkcolor=blue, citecolor=blue, urlcolor=blue} 

\graphicspath{{./Figs/}}

\newcommand{\beq}[1]{\begin{equation}\label{#1}}
\newcommand{\eep}{\;.\end{equation}}
\newcommand{\eec}{\;,\end{equation}}
\newcommand{\eeq}{\end{equation}}

\newcommand{\lb}{\left(}
\newcommand{\rb}{\right)}

\newcommand*\chem[1]{\ensuremath{\mathrm{#1}}} 

\newcommand{\g}{\gamma}

\newcommand{\G}{\Gamma}

\DeclareMathAlphabet{\mathcal}{OMS}{cmsy}{m}{n} 

\newcommand{\Hcal}{\mathcal{H}} 
\newcommand{\M}{\mathcal{M}}    

\newcommand{\bvec}[1]{\mathbf{#1}}

\newcommand{\kv}{\bvec{k}}
\newcommand{\av}{\bvec{a}}
\newcommand{\R}{\bvec{R}}
\newcommand{\rv}{\bvec{r}}

\newcommand{\HarvardPhysics}{Department of Physics, Harvard University, Cambridge, Massachusetts 02138, USA}
\newcommand{\HarvardSeas}{John A.~Paulson School of Engineering and Applied Sciences, Harvard University, Cambridge, Massachusetts 02138, USA}
\newcommand{\Rutgers}{
Department of Physics and Astronomy, Rutgers, The State University of New Jersey, Piscataway, NJ 08854, USA.}

\begin{document}

\title{
Stacking-dependent electronic structure of ultrathin perovskite bilayers
}

\author{Daniel T.~Larson}
\affiliation{\HarvardPhysics}
\author{Daniel Bennett}
\affiliation{\HarvardSeas}
\author{Abduhla Ali}
\affiliation{\HarvardSeas}
\affiliation{Cooper Union, New York, NY 10003, USA}
\author{Anderson S.~Chaves}
\affiliation{\HarvardSeas}
\author{Raagya Arora}
\affiliation{\HarvardSeas}
\author{Karin M. Rabe}
\affiliation{\Rutgers}
\author{Efthimios Kaxiras} \email{kaxiras@physics.harvard.edu}
\affiliation{\HarvardPhysics}
\affiliation{\HarvardSeas}

\begin{abstract}
Twistronics has received much attention as a new method to manipulate the properties of 2D van der Waals structures by introducing moir\'{e} patterns through a relative rotation between two layers.
Here we begin a theoretical exploration of twistronics beyond the realm of van der Waals materials by developing a first-principles description of the electronic structure and interlayer interactions of ultrathin perovskite bilayers.
We construct both an \emph{ab initio} tight-binding model as well as a minimal 3-band effective model for the valence bands of monolayers and bilayers of oxides derived from the Ruddlesden-Popper phase of perovskites, which is amenable to thin-layer formation.
We illustrate the approach with the specific example of Sr$_2$TiO$_4$ layers but also provide model parameters for Ca$_2$TiO$_4$ and Ba$_2$TiO$_4$.
\end{abstract}
\maketitle


\section{Introduction
\label{sec:intro}
}

Manipulating the properties of materials at the atomic scale has long been the goal of both physicists, for the purpose of exploring new physics, and materials scientists, for the purpose of creating new artificial structures with unusual properties that can improve the performance of various devices. 
This goal was advanced in a very substantial way with the  discovery of two-dimensional (2D) materials, starting with graphene, followed by the semiconducting families of transition-metal mono- and di-chalcogenides, insulators like hexagonal boron nitride, phosphorene, and a large number of other related structures \cite{GeimGrigorieva-2013}. 
Due to their atomic scale thickness, these materials represent the ultimate size limit in the out-of-plane direction.

More recently, an additional degree of freedom has been introduced to further manipulate the properties of 2D materials, namely the creation of moir\'{e} patterns in systems comprised of two or more stacked 2D layers, with length scales typically much larger than the original crystal lattice constants \cite{Andrei-2021}. 
The interplay of the atomic and moir\'{e} scales has proven very interesting as it allows for a wide range of emergent phenomena, including fractional \cite{xie2021fractional,zeng2023thermodynamic,park2023observation} and anomalous \cite{nuckolls2020strongly,wu2021chern} Hall effects, superconductivity \cite{cao2018unconventional,lu2019superconductors,chen2019signatures,park2021tunable,park2022robust}, correlated states \cite{cao2018correlated,Yazdani2023}, and the appearance and manipulation of polar \cite{yasuda2021stacking,bennett2022electrically,bennett2022theory,ko2023operando,bennett2023polar,van2024engineering,yasuda2024ultrafast} and magnetic order \cite{tong2018skyrmions,hejazi2020noncollinear,song2021direct,bennett2024stacking} in 2D. 
The creation of moir\'{e} patterns relies on carefully controlling the relative twist (referred to as ``twistronics'' \cite{carr2017twistronics}) or strain (lattice mismatch) in successive layers.

These effects have been demonstrated and explored in systems where the interaction between layers is of van der Waals (vdW) type, that is, rather weak and long-range compared to the interactions of the atoms {\em within} each layer, which are typically covalent or ionic.
This dichotomy in bonding within and across layers makes it possible to isolate single layers by exfoliation. 
From the single layers, it is then feasible to create stacked bilayer and multilayered assemblies using the tear-and-stack method \cite{yoo2019atomic}.

It is intriguing to consider the consequences of applying twistronics or strain in layered arrangements of other materials, such as oxides.
In contrast to vdW materials, where correlation effects often arise only as a direct consequence of the twisting, many oxide materials inherently exhibit strong correlation effects \cite{dagotto2005complexity}, which could then be tuned or enhanced by the fabrication of moir\'e superlattices.

\begin{figure}[t!]
\centering
\includegraphics[width=\columnwidth]{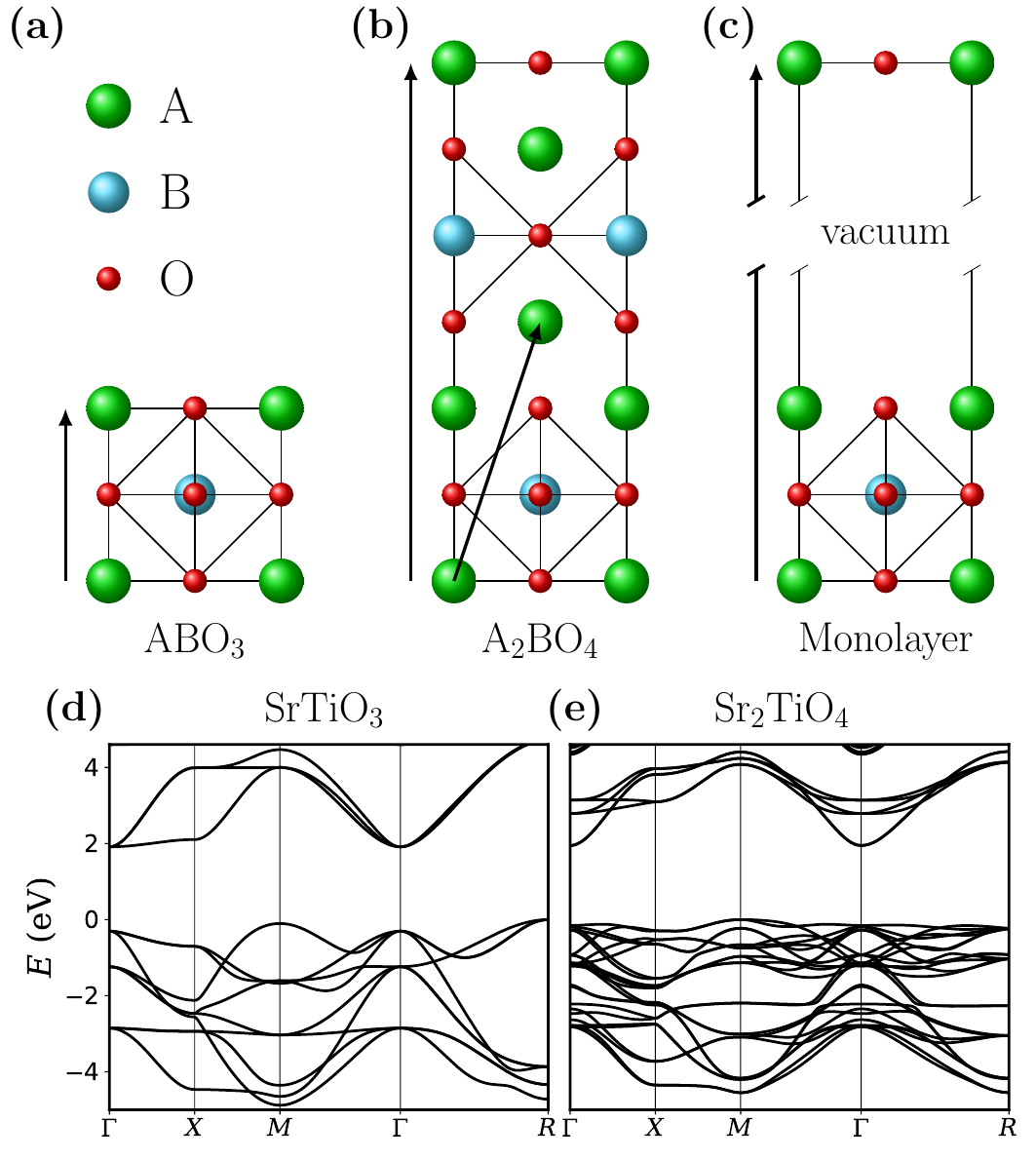}
\caption{
{\bf (a)} The unit cell of the \chem{ABO_3} bulk perovskite structure with the vertical lattice vector indicated.
Green spheres represent A atoms, blue spheres B atoms and red spheres O atoms.
{\bf (b)} A ``conventional" cell containing two units of the RP$_1$ phase \chem{A_2BO_4}, 
with the second unit shifted by the primitive lattice vector $\mathbf{a}_3 = (a/2, a/2, c)$. The vertical arrow indicates the periodic repetiton of the doubled unit cell. 
{\bf (c)}  An isolated monolayer constructed from the RP$_1$ unit cell.
{\bf (d)} Band structure of the bulk cubic \chem{SrTiO_3} crystal shown in \textbf{(a)}. {\bf (e)} Band structure of the bulk RP$_1$ \chem{Sr_2TiO_4} crystal shown in \textbf{(b)}. 
The zero of energy is set to the top of the valence band.
}
\label{fig:RP-structures}
\end{figure}

Many oxides crystallize in the perovskite structure with chemical composition per unit cell ABO$_3$ (Fig.~\ref{fig:RP-structures}(a)), where A is typically 
an atom from the first or second column of the Periodic Table and B is a transition metal.
In the cubic phase with periodic boundary conditions in all three Cartesian directions, the three O atoms in this unit form the corners of a regular octahedron.
This type of bulk structure does not easily lend itself to layered structures with stable surfaces, although perovskites have been fabricated in very thin films, a few unit cells in thickness, using molecular beam epitaxy \cite{ji2019freestanding}.

There are, however, variations on the bulk perovskite structure that contain natural cleavage planes, that is, planes along which the crystal can be easily separated in parts with stable surfaces. 
One example is the family of copper-oxides which exhibit high-temperature superconductivity, such as \chem{YBa_2Cu_3O_7} (YBCO) \cite{beno1987structure,schneemeyer1987superconductivity}, \chem{Bi_2Sr_2CaCu_2O_8} (BSCCO) \cite{massidda1988electronic,krishana1997plateaus}, and \chem{La_{2-x}Sr_{x}CuO_4} (LSCO) \cite{logvenov2009high}. 
Other examples are the so-called Ruddlesden-Popper phases \cite{RP-phase-1958}, with the chemical composition of the unit cell given by RP$_n$ = A$_{n+1}$B$_n$O$_{3n+1}$, which can be thought of as $n$ copies of the perovskite bulk unit terminated by an extra AO plane, with the next unit cell shifted by half of the diagonal of the in-plane unit cell. 
The RP$_1$ structure is shown in Fig.~\ref{fig:RP-structures}(b).
The presence of the extra AO plane introduces the possibility of creating stable layered forms of these materials \cite{schaak2000prying,stoumpos2016ruddlesden,lu2016synthesis,hong2020extreme,pan2021deterministic}.

Twisted perovskite structures have been fabricated using strontium titanate  (\chem{SrTiO_3}, STO) \cite{shen2022observation} and barium titanate  (\chem{BaTiO_3}, BTO) \cite{sanchez20242d} using a sacrifical-layer approach, as well as halide perovskites using other fabrication techniques \cite{wang2021colorful,zhang2023square}.
Twisting perovskites results in moir\'e superlattices with square-shaped domains, as opposed to the triangular or honeycomb domains formed by twisting the most common vdW layers.
First-principles calculations have recently been performed to characterize structural and electronic properties of twisted BTO layers \cite{lee2024moir}, including a phenomenological model of large-angle twisted structures. 

In this work, we construct \emph{ab initio} tight-binding models for RP$_1$ monolayers as well as bilayers in several high symmetry stacking configurations, in a similar manner to the minimal models developed for twisted vdW materials \cite{carr2019derivation,carr2019exact,bennett2024twisted}.
We complement the \emph{ab initio} models with simplified empirical tight-binding models describing only the three highest valence bands of RP$_1$ single layers. 
We provide parameters for the tight-binding models of three specific oxides: calcium titanate (CaTiO$_3$, CTO), STO, and BTO. 
These models serve as the foundation for continuum models of twisted oxide bilayers.

\section{Structural models of oxide layers}

We focus on RP$_n$ phases of perovskites as the prototypical structure of perovskite-based layers of atomic-scale thickness. 
For ${n\to\infty}$ this type of structure approaches the bulk perovskite structure.
For ${n=1}$, the basic unit is \chem{A_2BO_4}, whose crystal structure and the corresponding layered structure are shown in Fig.~\ref{fig:RP-structures}.

In the perovskite bulk structure shown in Fig.~\ref{fig:RP-structures}(a), the A atoms form a cubic lattice with lattice constant $a$: $\av_1 = a {\hat x}$, $\av_2 = a {\hat y}$, $\av_3 = a {\hat z}$.  
The cube with A atoms at its vertices has O atoms in the center of each face and a B atom in the center of the cube. 
In each unit cell the O atoms are arranged at the corners of a regular octahedron surrounding the B atom. 
In most perovskite oxides the bonds between atoms that make this structure stable are predominantly of covalent-ionic nature between the O atoms at the corners of the octahedron and the B atom at its center. 
The role of the A atoms is primarily to provide additional electrons that can be accommodated by the bonding states created by the interaction between B and O atoms. 
The right combination of A and B atoms in the bulk structure leads to all such bonding states being filled, with a large band gap separating them from the antibonding states, resulting in a stable insulating phase as shown in Fig.~\ref{fig:RP-structures}(d). 

In the RP$_1$ structure the presence of the extra plane of AO atoms leads to a natural division between structural units in the $\hat{z}$-direction, perpendicular to the basal $xy$-plane that contains the B atom.
Successive octahedra along the $\hat{z}$-axis are staggered by half of a unit-cell diagonal parallel to the basal plane, so as to 
give a rocksalt-like arrangement.

Considering a single \chem{A_2BO_4} unit separate from its neighbors along the $\hat{z}$ axis results in a layer, shown in Fig.~\ref{fig:RP-structures}(c), which is stable in the structural and electronic sense.
Specifically, in this single unit there are no broken bonds between the B atoms and the surrounding O atoms, and thus no dangling bonds on either surface of the layer. 
This is the thinnest stable structure that retains the essence of the perovskite parent but can still be thought of as a 2D layer that can potentially be twisted and stacked.
Thus we begin our study of moir\'{e} physics in the oxides with the properties of a single ``ultrathin" perovskite layer, followed by a study of bilayers.
For the latter we consider the elemental features and steps for describing their twistronics.

\section{Electronic structure of a single RP$_1$ oxide layer}

As in the case of vdW bilayers and multilayers, it is important to establish simple models that capture the essential physics of the single layer.
Such models, usually in the form of tight-binding hamiltonians with a basis consisting of the smallest possible number of atomic orbitals~\cite{fang2016electronic,fang2019angle}, can then serve to describe the low-energy physics in the moir\'{e} supercells in a clear and intuitive way.
A reliable method to produce such models is to use {\em ab initio} electronic structure calculations and construct localized Wannier orbitals \cite{mostofi2008wannier90} as the basis for the tight-binding hamiltonian. 
There is, however, considerable gauge freedom in the choice of Wannier orbitals. 
Therefore, a physically intuitive set of atomic orbitals must be used as a starting point, in order to derive the simplest and most useful set of localized orbitals for the basis.

\begin{figure}[!t]
\centering
\includegraphics[width=0.5\columnwidth]{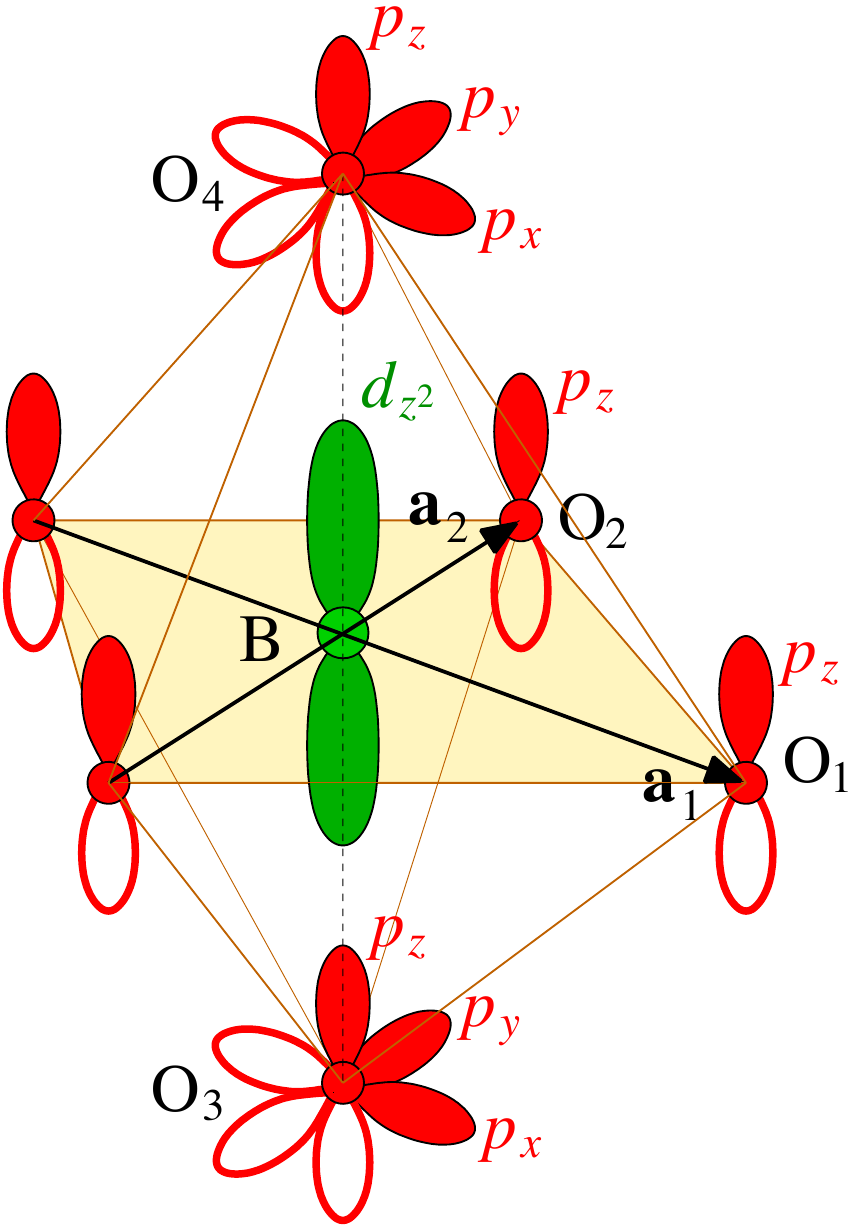}
\caption{
    The basic structural model, consisting of an octahedron of O atoms and a transition metal atom at its geometric center. 
    The O atoms on the mid-plane of the octahedron (highlighted in yellow) are labeled O$_1$ and O$_2$, and the two other O atoms at the apexes of the octahedron are labeled O$_3$ and O$_4$.
    The transition-metal atom at the geometric center is labeled B. The lattice vectors for the periodic arrangement of the
    O atoms on the mid-plane are shown and labeled ${\bf a}_1$, ${\bf a}_2$.  The $p_\ell$ (where $\ell = x,y,z$) atomic  orbitals of the O atoms are shown in red (filled and open  lobes correspond to positive and negative values), and the $d_{z^2}$  orbital of the B atom is shown in green (two positive lobes).
}
\label{fig:octahedron}
\end{figure}

\begin{figure*}
\centering
\includegraphics[width=\textwidth]{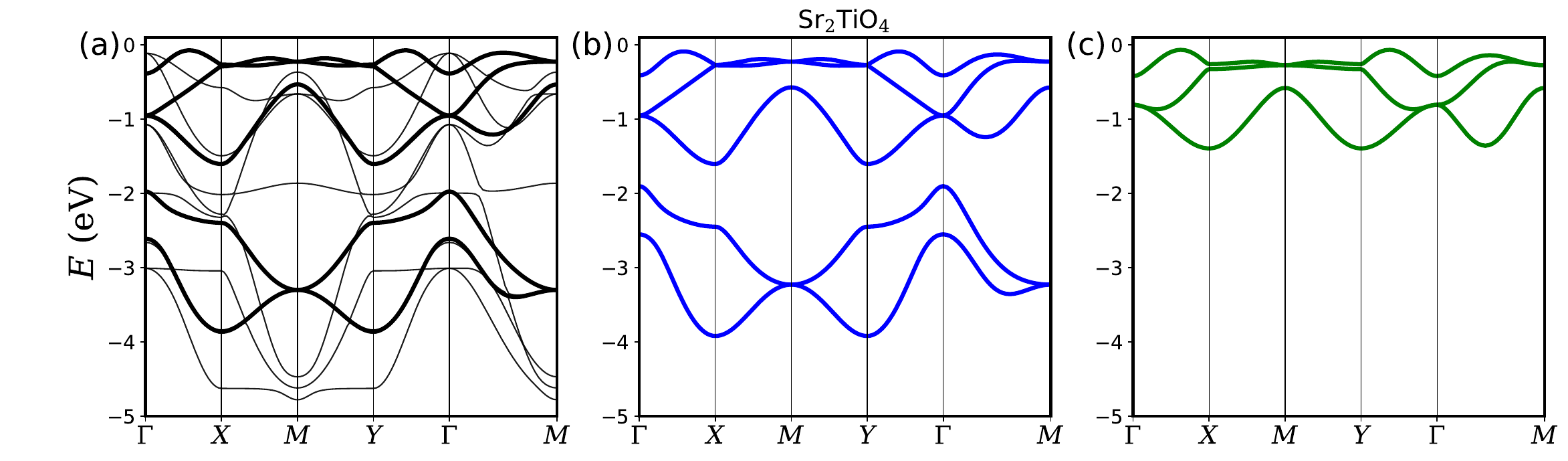}
\includegraphics[width=1\textwidth]{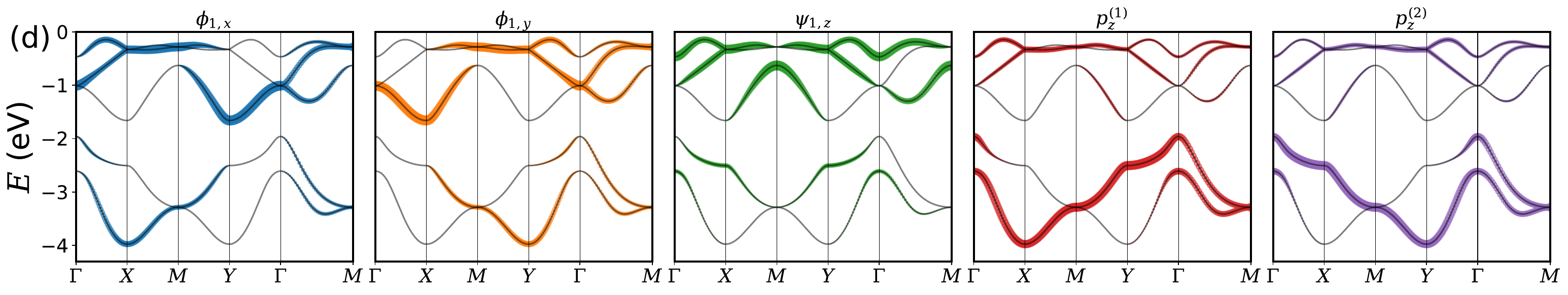}
\caption{STO RP$_1$ monolayer band structure: 
{\bf (a)} The 12 valence energy bands calculated using DFT are plotted with thin black lines, with the 5 relevant $\M$-odd bands highlighted in thicker black lines. 
The zero of the energy scale corresponds to the valence band maximum.
{\bf (b)} The 5 bands in the $\M$-odd sector described by Eqs.~\eqref{eq:H-tot}-\eqref{eq:Hc} using MLWF parameters from Table~\ref{tab:params}. 
{\bf (c)} The three bands obtained from the simplest model for describing the electronic structure of the STO RP$_1$ oxide unit, 
namely the hamiltonian in Eq.~\eqref{eq:H-tot} with $\Hcal^{(c)}_{\kv}$ set to zero and remaining parameters given in Table~\ref{tab:params-0}.
{\bf (d)} Orbital character of the 5-band MLWF tight-binding model.
}
\label{fig:5-bands_0}
\end{figure*}

Using {\em ab initio} density functional theory (DFT) calculations (see Appendix~\ref{sec:dft}), and physical intuition about the relevant atomic states involved in forming the stable oxide phases mentioned above, we propose the following basis: of the 7 atoms in the unit cell of the A$_2$BO$_4$ RP$_1$ slab,
the two A atoms mainly serve to donate their electrons to fill the valence bands and stabilize the structure; consequently their orbitals lie far from the Fermi level and do not contribute in any other important way to the electronic structure. 
That leaves 12 $p$-orbitals ($p_\ell$, $\ell=x,y,z$) associated with the four O atoms and the 5 $d$-orbitals of the B (transition metal) atom.
From those 17 orbitals we will select the valence states most important for interlayer interactions.
Five states, the $p_x$ and $p_y$ orbitals of the oxygen atoms in the mid-plane of the octahedron, and the $d_{x^2-y^2}$ orbital of the B atom, are involved in forming bonding, anti-bonding and non-bonding states that are localized at the mid-plane and therefore are expected to play a much reduced role in the interlayer coupling.
Three more orbitals, the $d_{xy}$, $d_{yz}$ and $d_{zx}$ orbitals of B lie much higher in energy and form the bottom of the conduction band, separated by a large band gap of several eV from the top of the valence band. 
This leaves only 9 orbitals potentially involved in the interactions between layers, namely the $p_x$ and $p_y$ orbitals of the oxygen atoms located on the top and bottom surfaces, the $p_z$ orbitals of all four O atoms and the $d_{z^2}$ orbital of atom B, all of which are shown in Fig.~\ref{fig:octahedron}.
We construct from those 9 orbitals linear combinations which represent the appropriate basis of the minimal tight-binding hamiltonian. 
These symmetry adapted states are divided in three groups: 
First, we note that the crystal structure of the layer is symmetric under a mirror reflection $\M$ about the mid-plane of the octahedron, with $\M^2 = I$. Under this symmetry, the orbitals of interest transform as follows:
\begin{align}
\M p_z^{(1)} & = -p_z^{(1)}, \quad
\M p_z^{(2)} = -p_z^{(2)}, \quad
\M p_z^{(3)} = -p_z^{(4)}, \\ 
\M p_x^{(3)} &= p_x^{(4)}, \quad
\M p_y^{(3)} = p_y^{(4)}, \quad
\M d_{z^2} = d_{z^2}
\end{align}
where the superscript identifies the O atom to which the orbital belongs, according to the scheme shown in Fig.~\ref{fig:octahedron}.
Accordingly, we choose the linear combinations 
that possess odd or even symmetry according to this 
 mirror operation and label the odd ones by odd integers 
($\pm 1$) and the even ones by even 
integers ($0$,$2$) in the following; 
orbitals of different symmetry do not couple. 
An important group of states is the linear combinations 
of the $d_{z^2}$ orbital of B (which is even under $\M$) with the $p_z$ orbitals 
of the apical O atoms, denoted as:
\begin{align}
\psi_{0,z}  &=  
\frac{1}{\sqrt{2}} d_{z^2} +  \frac{1}{2} \left( p_z^{(3)} 
- p_z^{(4)} \right ), 
\label{eq:psi_orbitals-1}\\
\psi_{1,z}  &=  \frac{1}{\sqrt{2}}
\left (p_z^{(3)} + p_z^{(4)} \right ),
\label{eq:psi_orbitals-3}\\
\psi_{2,z}  &= 
\frac{1}{\sqrt{2}} d_{z^2} -  \frac{1}{2} \left( p_z^{(3)} 
- p_z^{(4)} \right ). 
\label{eq:psi_orbitals-2}
\end{align}
The first of these, $\psi_{0,z}$, is a bonding combination 
and gives rise to a low-energy state that is fully occupied.
The third, $\psi_{2,z}$, is an anti-bonding combination 
and gives rise to a high-energy state which is unoccupied. 
Finally, $\psi_{1,z}$ is a non-bonding state which 
is odd under $\M$ and thus doesn't couple to $d_{z^2}$.

Another group of states consists of linear combinations 
of the $p_x$ and $p_y$ orbitals of the apical O atoms, 
namely
\begin{align}
\phi_{0,\ell} &= \frac{1}{\sqrt{2}} \left ( 
p_\ell^{(3)} + p_\ell^{(4)} \right ), \quad \ell = x,y,
\label{eq:phi_orbitals-1}
\\
\phi_{1,\ell} &= \frac{1}{\sqrt{2}} \left ( 
p_\ell^{(3)} - p_\ell^{(4)} \right ), \quad \ell = x,y,
\label{eq:phi_orbitals-2}
\end{align}
the first even and the second odd under $\M$.

Finally, the remaining two possible states are the 
linear combinations 
of the oxygen $p_z$ orbitals that lie in the mid-plane, namely
\begin{align}
\chi_{1,z} & = \frac{1}{\sqrt{2}} \left ( 
p_z^{(1)} + p_z^{(2)} \right ),
\label{eq:chi_orbitals-1}\\
\chi_{-1,z} & = \frac{1}{\sqrt{2}} \left ( 
p_z^{(1)} - p_z^{(2)} \right ) 
\label{eq:chi_orbitals-2}
\end{align}

Because we are focusing on states that will be involved in interlayer interactions, the basis of mutually orthogonal hybrid orbitals 
defined in Eqs~\eqref{eq:psi_orbitals-1}--\eqref{eq:chi_orbitals-2} can be further reduced by considering the following:

\begin{itemize}
\item The state $\psi_{2,z}$ is an anti-bonding state whose energy lies well above the conduction band minimum, and can thus be removed from further consideration.
\item The bonding state $\psi_{0,z}$ has very low energy, as expected due to its character.
Thus it can also be removed from the basis of relevant valence states.
\item The pair of states $\phi_{0,\ell}$ $(\ell=x,y)$ consist
of O $p_x$ and $p_y$ orbitals whose wavefunctions do not extend much outside the slab and are less affected 
by interlayer interactions. 
Being $\M$ even, they also do not couple to the remaining odd valence states. 
These two states can also be neglected from further consideration.
\end{itemize}

The preceding analysis suggests that only five $\M$-odd states are relevant to the physics of interlayer interactions, namely 
$\psi_{1,z}$, 
$\phi_{1,\ell}$ $(\ell=x,y)$, and $\chi_{\pm 1,z}$. 
The {\em ab initio} calculations for \chem{Sr_2TiO_4} (which we also refer to as STO) shown in Fig.~\ref{fig:5-bands_0}(a) support this conclusion, since the top of the valence band is almost exclusively derived from the $\M$-odd states.
Based on the DFT calculations of bilayers, we find that the valence band maximum arises from $\mathcal{M}$-odd states that are shifted to higher energies due to interlayer interactions, suggesting that it is reasonable to ignore the 
$\M$-even bands that produce the highest energy valence states at $\G$.
To model the 5 bands of the $\M$-odd sector we use a basis of maximally localized Wannier functions (MLWF) derived from first-principles calculations (for details see Appendix~\ref{sec:dft}).
We construct the following tight-binding hamiltonian that retains the most important hopping terms needed to reproduce the band structures for the RP$_1$ monolayer.
In the basis $(\phi_{1,x}, \phi_{1,y}, \psi_{1,z}, p_z^{(1)}, p_z^{(2)})$ the hamiltonian is given in block form by:
\beq{eq:H-tot}
 \Hcal_{\kv} = \begin{bmatrix}
     \Hcal^{(3)}_{\kv} & \Hcal^{(c)}_{\kv} \\
  h.c. & \Hcal^{(2)}_{\kv}
 \end{bmatrix}
\eec
where $h.c.$ stands for hermitian conjugate 
and the blocks are given by
\begin{widetext}
\beq{eq:H3}
\Hcal_{\kv}^{(3)} = \begin{bmatrix}
\epsilon_{xy} + t_{xy} h_{\kv}(\av_1,\av_2,\lambda_{xy},\mu_{xy}) & 
t'_{xy} g_{\kv}(\av_1)g_{\kv}(\av_2) & 
t g_{\kv}(\av_1) \\
& 
\epsilon_{xy} + t_{xy} h_{\kv}(\av_2,\av_1,\lambda_{xy},\mu_{xy}) & 
t g_{\kv}(\av_2) \\
h.c. &  
& 
\epsilon_z + t_z h_{\kv}(\av_1,\av_2,1,\mu_z)
\end{bmatrix}
\eec
\end{widetext}
\begin{align}
\label{eq:H2}
\Hcal_{\kv}^{(2)}& = \begin{bmatrix}
\epsilon_{0} + t_{0} f_{\kv}(\av_1) & t'_0 f_{\kv}(\frac{\av_1}{2}) f_{\kv}(\frac{\av_2}{2}) \\
h.c. & \epsilon_{0} + t_{0} f_{\kv}(\av_2)
\end{bmatrix},
\\
\label{eq:Hc}
\Hcal_{\kv}^{(c)} &= \begin{bmatrix}
    t_2 j^{(2)}_{\kv}(\av_1,\av_2,\nu_2,\nu_2') & 0 \\
0 & t_2 j^{(2)}_{\kv}(\av_2,\av_1,\nu_2,\nu_2') \\
t_1 j^{(1)}_{\kv}(\av_1,\av_2,\nu_{1}) 
& t_1 j^{(1)}_{\kv}(\av_2,\av_1,\nu_{1})  \\
\end{bmatrix}.
\end{align}
The momentum dependence in $\Hcal_{\kv}$ is described by the functions $f_{\kv}$ and $g_{\kv}$\footnote{The arguments to these functions, such as $\av_1$, $\av_2$, $\mu$, $\nu$, etc., are understood as discrete parameters, with $\kv$ the only independent variable.}:
\begin{align}
    f_{\kv}(\av) &= e^{i\kv\cdot\av} + e^{-i\kv\cdot\av} = 2\cos(\kv\cdot\av) \label{eq:fk-definition} \\ 
    g_{\kv}(\av) &= e^{i\kv\cdot\av} - e^{-i\kv\cdot\av} = 2i\sin(\kv\cdot\av),
    \label{eq:gk-definition}
\end{align}
as well as the combinations:
\begin{equation}
    h_{\kv}(\av,\av',\lambda,\mu)  = f_{\kv}(\av) + \lambda f_{\kv}(\av') 
    + \mu f_{\kv}(\av) f_{\kv}(\av') 
    \label{eq:hk-definition}
\end{equation}
\begin{align}
    j^{(1)}_{\kv}(\av,\av',\nu)  & = \left[ 1 + \nu f_{\kv}(\av')\right ] f_{\kv}\left(\frac{\av}{2}\right) \\
    j^{(2)}_{\kv}(\av,\av',\nu,\nu')  & = \left[ 1 + \nu f_{\kv}(\av')\right ] g_{\kv}\left(\frac{\av}{2}\right)
    + \nu' g_{\kv}\left (\frac{3\av}{2}\right).
    \label{eq:j1-j2-definition}
\end{align}
The values of the parameters involved are collected in Table~\ref{tab:params} for three representative RP$_1$ cases, namely CTO, STO and BTO. 
For STO, Fig.~\ref{fig:5-bands_0}(b) demonstrates that the $\M$-odd bands of the 15-parameter MLWF TB model are nearly indistinguishable from the bands obtained through DFT calculations shown in Fig.~\ref{fig:5-bands_0}(a). 
The same is true for the band structures of CTO and BTO, which are shown in Supplemental Figs.~S4 and S5.
All of the parameters in Table~\ref{tab:params} can be interpreted as arising from hopping between $p$-orbitals at the oxygen sites, modified slightly by interactions with the Ti atom.
For instance, $\epsilon_{xy}$ and $\epsilon_z$ give the onsite energies of the $\mathcal{M}$-odd $p_x$, $p_y$, and $p_z$ orbital combinations located at O$_3$ and O$_4$.
Similarly, $t_{xy}$ and $\bar{t}_{xy}$ give first and second neighbor hoppings from $p_x$ to itself, $t'_{xy}$ arises from $p_x$-$p_y$ hopping along a diagonal (2nd neighbor hopping), and $t$ arises from $p_x$-$p_z$ hopping.
This latter coupling would vanish for $p$-orbitals in a single plane, but the $\mathcal{M}$-odd combinations allow for nonzero matrix elements between $p_x$ and $p_z$ orbitals on the top and bottom surfaces, respectively.
The remaining parameters represent additional onsite and farther neighbor interactions, many of which are related to each other by the crystal symmetry.
\begin{table*}[ht!]
  \centering
  \caption{Values of the parameters of the tight-binding hamiltonian in Eqs~\eqref{eq:H-tot}-\eqref{eq:Hc} for models of CTO, STO, and BTO.
  The relations between values listed here and the parameters that appear in Eqs~\eqref{eq:H3}-\eqref{eq:Hc} are:
  $\mu_{xy} = {\bar t}_{xy}/t_{xy}$, $\nu_1={\bar t}_1/t_1$, $\nu_2 = {\bar t}_2/t_2$, $\nu_2'=t_2'/t_2$;
a satisfactory agreement with the first-principles results is obtained by setting $\mu_z=0$, thus reducing the number of physical parameters needed to 15. 
All values are in eV except $\lambda_{xy}$ which is dimensionless. Numbers in parentheses indicate the most important changes to these intralayer parameters when in an AB-stacked bilayer (see Sect.~\ref{sec:interlayer} for details).
}
  \begin{tabular}{|l|r|r|r|r|r|r|r|r|r|r|r|r|r|r|r|}
  \hline
   & $\epsilon_{xy}-\epsilon_z$ & $\epsilon_{0}-\epsilon_z$ & $t_{xy}$ & $\lambda_{xy}$ & $\bar{t}_{xy}$ & $t'_{xy}$ & $t_z$  & $t_{0}$ & $t'_0$  & $t$ & $t_2$ & ${\bar t}_2$ & $t_2'$ & $t_1$ & ${\bar t}_1$ \\ \hline
    CTO & $-0.795$ & $-2.185$ & $0.027$ & $-0.111$ & $0.058$ & $0.054$ & $-0.025$ & $0.150$  & $-0.076$ & $0.074$ & $0.867$ & $0.051$ & $0.044$ & $0.352$ & $-0.025$ \\
    & $(-0.576)$ & $(-1.749)$ & & & & & ($-0.014$) & & & & (0.839) & & & & \\ \hline
   STO & $-0.596$ & $-1.714$ & $-0.021$ & $0.524$ & $0.087$ & $0.079$ & $-0.017$ & $0.138$  & $-0.044$ & $0.067$ & $0.772$ & $0.068$ & $0.037$ & $0.335$ & $-0.037$ \\ 
    & $(-0.467)$ & $(-1.440)$ & & & & & $(-0.010)$ & & & & $(0.748)$ & & & & \\ \hline
     BTO & $-0.443$ & $-1.141$ & $-0.100$ & $0.190$ & $0.126$ & $0.115$ & $-0.002$ & $0.107$ & $0.009$ & $0.049$ & $0.652$ & $0.082$ & $0.023$ & $0.315$ & $-0.057$ \\
      & $(-0.356)$ & $(-0.977)$ & & & & & $(0.004)$ & & & & $(0.627)$ & & & & \\
    \hline
  \end{tabular}
  \label{tab:params}
\end{table*} 

It is useful to reduce the number of bands to the absolute minimum needed to capture the low-energy physics in a bilayer system.
It is clear that the 5 $\M$-odd bands separate 
into two sectors: one sector in the lower half of the valence band manifold, composed primarily of the states 
$\chi_{\pm 1,z}$, and another sector closer to the valence band maximum (VBM), 
consisting of combinations
of $\psi_{1,z}$ and $\phi_{1,\ell}$ $(\ell=x,y)$, as shown in the plot of orbital character in Fig.~\ref{fig:5-bands_0}(d).
While mixing between all 5 states is allowed by the $\M$ symmetry, our goal is to ``integrate out" the lower two states and build a simplified 3-band model that describes the top of the valence band and contains the states most relevant to any interlayer interactions.
The separation into two subsets of bands is reflected in the values of the corresponding on-site energy terms; for all three materials $\epsilon_{xy}-\epsilon_z$ is roughly one third of $\epsilon_0-\epsilon_z$, demonstrating that the first three states in the basis lie closer to the VBM.
The lower two bands of the $\M$-odd sector can be formally integrated out based on the Schur complement, which leads to the following eigenvalue problem for the top three bands (represented by $\Psi$):
\begin{equation} \label{eq:schur}
    \left[ \Hcal_{\kv}^{(3)} + \Hcal_{\kv}^{(c)} \left(E - \Hcal_{\kv}^{(2)} \right)^{-1} \Hcal_{\kv}^{(c)\dagger} \right] \Psi 
    = E \Psi.
\end{equation}
To the extent that bands arising from 
$\Hcal_{\kv}^{(2)}$ are at a much lower energy and their dispersion can be 
neglected, they can be represented by a single average energy scale $\epsilon$, so the additional contributions to 
$\Hcal_{\kv}^{(3)}$ (second term in square brackets in Eq.\eqref{eq:schur}) can be approximated by
\begin{equation}
    \delta\Hcal_{\kv}^{(3)} \sim \frac{1}{\epsilon} \Hcal_{\kv}^{(c)}\Hcal_{\kv}^{(c) \dagger }.
\end{equation}
This correction modifies the numerical values of the parameters appearing in $\Hcal_{\kv}^{(3)}$ and
introduces additional terms representing hopping to further neighbors.
Intuitively, these additional hopping terms can be understood as being mediated by the coupling to $p_z$ orbitals on O atoms in the mid-plane  (O$_1$ and O$_2$ in Fig.~\ref{fig:octahedron}), since the hamiltonian  $\Hcal_{\kv}^{(3)}$ acts only on the orbitals $\psi_{1,z}$ and $\phi_{1,\ell}$ ($\ell = x,y$) located on the apical O atoms (O$_3$ and O$_4$ in Fig.~\ref{fig:octahedron}).
The significant coupling between orbitals on the O$_{1,2}$ and O$_{3,4}$ atoms, along with the moderate energy scale of the $\chi_{\pm1,z}$ states, is what makes the further neighbor hopping terms important for the simplified model.
This coupling is also evident in the orbital character of the bands shown in Fig.~\ref{fig:5-bands_0}(d), where the states $\psi_{1,z}$ and $\phi_{1,\ell}$ ($\ell = x,y$) have projections predominantly in the top three bands, but also non-negligible projections in the lower two bands. 
Conversely, the states $p_z^{(1,2)}$ have projections predominantly in the lower two bands but also to a lesser extent in the top three bands. 
Rather than using Eq.~\eqref{eq:schur} directly, for simplicity we numerically fit the energies of the three $\M$-odd states at the top of the valence band using the hamiltonian $\Hcal_{\kv}^{(3)}$, which involves just 7 parameters.
The values of these parameters are given in Table~\ref{tab:params-0} and the resulting bands are shown in Fig.~\ref{fig:5-bands_0}(c). 
The differences between the parameters in this simplified model and the MLWF version can be understood as the average renormalization of the parameters due to interactions 
between orbitals on the apical O atoms and 
those on O atoms at the mid-plane.
The three-band model captures the qualitative behavior of the bands with a minimal number of parameters.
\begin{table}[!ht]
  \caption{Values of the 
  parameters (in eV) of the tight-binding hamiltonian in Eq.~(\ref{eq:H3}) for models of CTO, STO, and BTO.
  The relations between values listed here 
  and the parameters that appear in 
   Eq.~\eqref{eq:H3} are:
  $\mu_{xy} = {\bar t}_{xy}/t_{xy}$, 
  $\mu_{z} = {\bar t}_{z}/t_{z}$. 
  For these fits we have set 
  $\lambda_{xy} = -1/3$, which leaves 
  7 adjustable parameters, since only differences between onsite energies $\epsilon$ are physically meaningful. Numbers in parentheses indicate the most important changes to these intralayer parameters when in an AB-stacked bilayer (see Sect.~\ref{sec:interlayer} for details).
  }
  \begin{tabular}{|l|r|r|r|r|r|r|r|}
  \hline
   & $\epsilon_{xy}-\epsilon_z$  & $t_{xy}$ & $\bar{t}_{xy}$ & $t_{xy}'$ & $t_z$ & $t$ & $\bar{t}_z$ \\ \hline
  CTO & $-0.41$ & $-0.19$ & $0.03$ & $0.09$ & $0.01$ & $0.19$ & $-0.01$\\
  & $(-0.28)$ &  &  &  & & & \\ \hline
  STO & $-0.32$ & $-0.20$ & $0.04$ & $0.10$ & $0.02$ & $0.17$ & $-0.03$\\
  & $(-0.20)$ &  &  &  & & & \\ \hline
  BTO & $-0.39$ & $-0.30$ & $0.07$ & $0.13$ & $0.00$ & $0.11$ & $-0.02$\\   & $(-0.39)$ &  &  &  & & & \\ \hline
\end{tabular}
  \label{tab:params-0}
\end{table} 

\section{Ultrathin oxide bilayers}

Ultimately we aim to model a bilayer composed of two RP$_1$ monolayers described above, with a relative twist between them. 
Detailed calculations based on atomistic models rely on the construction of commensurate cells that can only host special twist angles.
In contrast to this, continuum models can be used to analyze the electronic structure for arbitrary (possibly incommensurate) twist angles~\cite{Bistritzer2011}. 
In the following we construct the foundation for tight-binding and continuum models
by first considering the local stacking arrangements generated by twisting square lattices, then modeling those arrangements as relative shifts between aligned bilayers, and finally discussing the full twisted structure.

\subsection{Geometric effects of layer twist}

\begin{figure}[t!]
\centering
\includegraphics[width=\columnwidth]{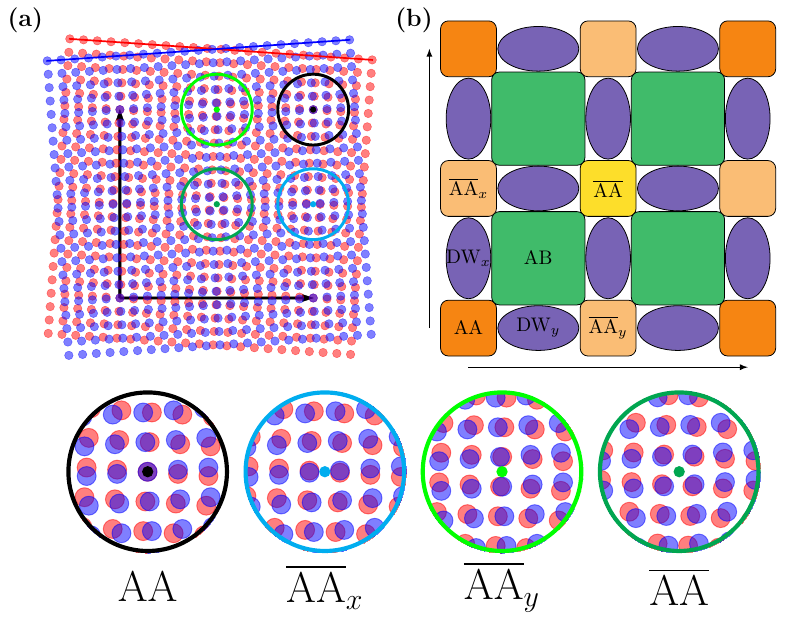}
\caption{
Illustration of the moir\'{e} pattern formed by two square 
lattices with a relative twist. 
{\bf (a)} A specific instantiation with atomic positions in the two lattices represented by blue and red dots, for a commensurate twist angle of $\theta = 8.17^{\circ}$;
the blue and red lines passing through the top rows of atoms intersect at angle $\theta$. 
The areas of closely matched lattices are shown by the colored circles, with magnified images labeled below. The colored dots at the center of the circles mark the coincidence of equivalent points from the primitive unit cells the two layers. 
Black arrows indicate the commensurate unit cell, which has a lattice constant that is twice the moir\'{e} length.
{\bf (b)} A schematic representation of the different types of domains, including the four AA-type stackings (in shades of orange) the AB stacking (in green) and the two types of domain walls DW$_x$, DW$_y$ (in purple).
}
\label{fig:6-moire}
\end{figure}

An arbitrary twist angle generally leads to an incommensurate structure which is ordered, but not periodic, on the atomic scale.
We first address the issue of the possible local environments that electrons experience at the interface of a twisted bilayer, in the spirit of the configuration space approach~\cite{carr2017twistronics, massatt2017electronic, cazeaux2020atomistic}.
The moir\'{e} pattern created by a relative twist between two square lattices consists of several different regions, characterized by the relative stacking between the layers.
One region, in which the atomic positions of the two layers are very closely matched in-plane and separated only in the out-of-plane direction, is referred to as an AA stacking region. 
This region is most easily identified visually because, when viewed along the out-of-plane direction, the close matching of the atomic positions leaves plenty of open space between atoms (circled regions in Fig.~\ref{fig:6-moire}(a)). 
Next, interspersed evenly between AA regions are regions in which atoms from the two layers are maximally mismatched in-plane.
Such regions are referred to as AB regions, and correspond to the bulk RP$_1$ stacking arrangement. 
Finally, the regions separating neighboring AA and AB regions are saddle points usually referred to as ``domain walls'' (DW).

Fig.~\ref{fig:6-moire}(a) shows a simplified diagram of two square lattices, each with a single ``atom" at every lattice site, and a relative twist between the layers. 
A regular pattern of AA-like regions forms; close inspection reveals that they are similar to each other, though not identical.
For some of the regions, referred to simply as ``AA" in the following, the center of the region has two atomic positions vertically aligned.
At the center of other AA-like regions it is the unit cell centers (devoid of atoms) that are vertically aligned. 
These are designated as $\overline{\rm AA}$.
There exist further variations on these, with other high-symmetry points in each of the two layers aligning at the center of an AA-type region. 
In the case of the square lattice, two such important variations are those in which the midpoint of the unit cell edge in the $x$-direction (referred to as $\overline{\rm AA}_x$ region) or along the $y$-direction (referred to as $\overline{\rm AA}_y$ region) are vertically aligned.
The four distinct AA stacking regions are labeled in Fig.~\ref{fig:6-moire}(b).

The conclusion from the preceding discussion is that for two square lattices a generic commensurate supercell contains four regions of AA-type stackings and four regions of AB-type stackings.
The four AA-type stackings are distinct, and only two, $\overline{\rm AA}_x$ and  $\overline{\rm AA}_y$, are related by the $C_{4z}$ symmetry. 
The four regions of AB-stacking are related by $C_{4z}$ symmetry; in other words, there is only one type of AB stacking. 
Finally, there are two types of DW stacking, namely DW$_x$ and DW$_y$, which are also related by $C_{4z}$ symmetry.
For incommensurate cells the distinctions between the regions and the symmetries that relate them are approximate. 
In the following we will assume the 4 different AA-type stackings have very similar effects on the electronic properties and consider them essentially equivalent, referring to them generically as AA stacking. 
As a result, the commensurate supercell is 4 times larger than the apparent moir\'{e} cell, or equivalently, the apparent moir\'{e} length scale is half of the commensurate supercell lattice constant, which is indicated by the black arrows in Fig.~\ref{fig:6-moire}
\footnote{For half of the commensurate angles a smaller supercell can be formed which is rotated by 45$^\circ$ and contains only two distinct AA regions.}.

\subsection{Stacking-dependent structural properties}

\begin{figure}[t!]
\centering
\includegraphics[width=0.8\columnwidth]{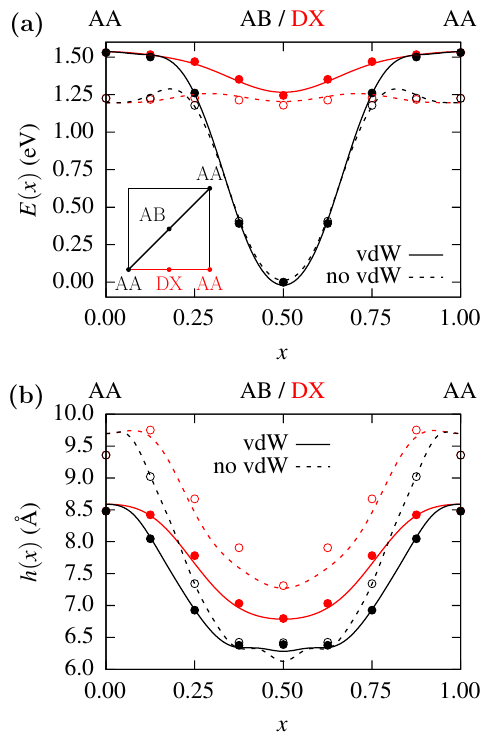}
\caption{
{\bf (a)} GSFE and 
{\bf (b)} GSFH line cuts for two STO RP$_1$ monolayers based on DFT calculations both with vdW corrections (solid symbols) and without (open symbols).
The DFT results were fit to smooth functions as described in Appendix~\ref{sec:appendix-GSFE}, which are plotted as solid and dashed lines.
}
\label{fig:test-vdw}
\end{figure}

We next investigate the properties of the RP$_1$ monolayers when they are stacked, without a twist, in the various configurations just identified. 
For two such layers we can calculate the generalized stacking fault energy (GSFE) and height (GSFH) \cite{kaxiras1993free} using first-principles calculations. 
These are the cohesive energy and interlayer separation, respectively, as a function of $\rv$, 
the two-dimensional vector that defines
the relative shift of one layer relative to the other.
The GSFE and GSFH for bilayer \chem{Sr_2TiO_4} are shown in Fig.~\ref{fig:2-GSFE}.
Both are even, periodic functions that can be easily fit with low-order Fourier harmonics (see Appendix~\ref{sec:appendix-GSFE}). Line cuts from Fig.~\ref{fig:2-GSFE} along high symmetry directions, along with the corresponding fits, are shown in Fig.~\ref{fig:test-vdw}.
As expected, the lowest energy configuration in the GSFE corresponds to the AB stacking of the layers, consistent with the stacking of successive units in the bulk RP$_1$ structure, and the highest energy configuration corresponds to AA stacking with the bottom apical O atom of the top-layer octahedron directly above the top apical O atom of the bottom-layer octahedron. 
Interestingly, the energy of the AA structure is more than 1 eV/unit cell higher than that of the AB structure. 
Moreover, in the AA structure the vertical distance between the top and bottom O octahedral apex atoms is quite large, about 4.4~\AA{}, which is almost a factor of 2 larger than the vertical separation of 2.3~\AA{} between the same atoms in the AB structure.

This dramatic difference in relaxed interlayer separation indicates that the interlayer interactions between the two layers are maximized in the AB configuration and minimized in the AA configuration. 
In the latter configuration the electronic states in the two layers are nearly non-interacting, as the electronic structure calculations discussed in the following section confirm. 
\begin{figure}[t!]
\centering
\includegraphics[width=\columnwidth]{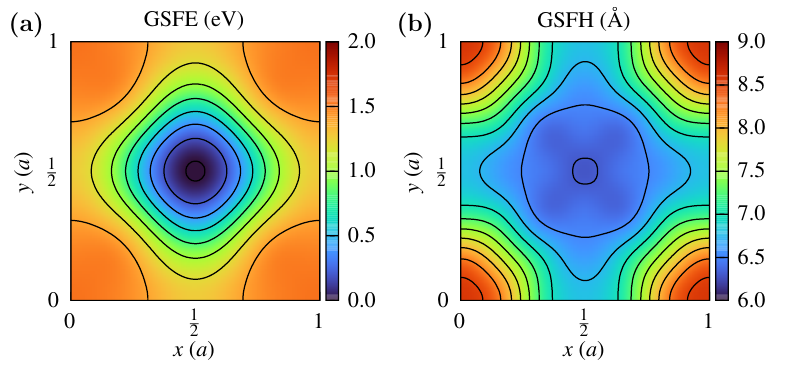}
\caption{
{\bf (a)} GSFE and 
{\bf (b)} GSFH for a STO RP$_1$ bilayer.
Results were obtained from first-principles calculations, including a vdW dispersion correction and allowing for optimization of the distance between the layers.
The high symmetry stackings are $\rv_{\rm AA} = \rv_{00}=  (0,0)$, $\rv_{\rm AB} = \rv_{11} = (\frac{1}{2},\frac{1}{2})$, $\rv_{{\rm DW}_{x}} = \rv_{10} =  (\frac{1}{2},0)$ and $\rv_{{\rm DW}_{y}} = \rv_{01} = (0,\frac{1}{2})$, in units of the lattice constant $a$.
}
\label{fig:2-GSFE}
\end{figure}

\subsection{Interlayer coupling}
\label{sec:interlayer}

Our next goal is to outline the derivation 
of the interlayer coupling matrix elements of the 
hamiltonian in the framework of the configuration space 
approach. In this framework, a moir\'{e} bilayer is 
described in terms of all the local atomic environments that arise due to the twist between the two layers.
These local environments can be approximated by considering 
the interface between two \emph{untwisted} layers that have a horizontal
shift $\rv$ of one layer relative to the other. 
For consistency, we consider the lower layer to be 
fixed and the upper layer to be shifted by $\rv$.
The hamiltonian describing such a bilayer
can be written in block form,
\begin{equation}
\label{eq:Hbilayer}
\mathcal{H}^\mathrm{bi}_\kv (\rv) = \begin{bmatrix}
    \mathcal{H}^\mathrm{mono}_\kv & \Hcal_{\kv}^{\rm int}(\rv) \\
    h.c. & \mathcal{H}^\mathrm{mono}_\kv
\end{bmatrix},    
\end{equation}
where $\mathcal{H}^\mathrm{mono}_\kv$ is the monolayer hamiltonian from either Eq.~\eqref{eq:H-tot} or Eq.~\eqref{eq:H3}.
The $\rv$ dependence of the intralayer blocks $\Hcal_\kv^\mathrm{mono}$  is expected to be quite small in comparison to the interlayer dependence.
However, in AB stacking the two layers are closest together and have the strongest interactions, modifying some of the monolayer parameters.
The intralayer parameters appropriate for AB stacking are given in parentheses in Tables~\ref{tab:params} and \ref{tab:params-0}.

We now describe $\Hcal_\kv^\mathrm{int}(\rv)$, beginning with some definitions that will facilitate notation.
In the configuration space framework, $\rv$ can take any value within the primitive unit cell, though many values are equivalent due to the crystal symmetry.
We take $\rv=\mathbf{0}\equiv \rv_{00}$ to be the AA stacking configuration, where the outer O atoms of the two layers, namely the upper apex O of the bottom layer (labeled O$_4$ in Fig.~\ref{fig:octahedron}) and the lower apex O of the top layer (labeled O$_3$ in Fig.~\ref{fig:octahedron}) are vertically aligned.
The DW$_x$/DW$_y$ stackings are at $\rv = \frac{1}{2}\av_{1}\equiv \rv_{10}$ and $\rv = \frac{1}{2}\av_{2}\equiv \rv_{01}$, respectively, and the AB stacking is $\rv =\frac{1}{2}\lb \av_{1} + \av_{2}\rb \equiv \rv_{11}$.
For a general stacking shift $\rv = (x,y)$, the hoppings between oxygen $p_{\ell}$ ($\ell=x,y,z$) orbitals depend on three quantities: the vertical interlayer separation, the in-plane distance between the O atoms in neighboring layers, and the angle between the axes of the $p_{\ell}$ orbitals of the different neighboring atoms. 
The relative shift vector $\rv$ gives the in-plane 
position of O$_3$ of the top layer. 
We denote the position of this atom along the direction 
perpendicular to the interface as $z$.

We denote the in-plane position of atoms in the lower layer by $\R_j$ and their position on the $z$ axis 
as $z_j$.
The index $j$ runs over all the atoms 
in the lower layer within hopping range of the 
atom O$_3$ in the top layer; 
this range is defined by a cutoff function $f_c(x)$
which is equal to 1 for $x<1$ and goes smoothly 
to zero for $x\geq 1$.
The other quantities that enter in the 
expressions for the interlayer hopping matrix elements 
are defined by:
\begin{align}
    \rv_j &= \R_j - \rv, \quad r_j = |\rv_j|, \quad 
    {\bar r}_j = \frac{r_j}{a}, 
    \label{eq:tofr_general-1}\\
    \delta {\bar z}_j &= \frac{|z_j - z|}{h_0}, \quad
        x_j = -\frac{\rv_j \cdot {\hat x}}{r_j}, \quad 
        y_j = -\frac{\rv_j \cdot {\hat y}}{r_j},
    \label{eq:tofr_general-2}
\end{align}
where $h_0$ is the height difference between the 
aligned positions (for $\rv=\mathbf{0}$) 
of top-layer atom O$_3$ and bottom-layer atom O$_4$.
To avoid any divergent behavior, 
which might arise in the expressions for $x_j$, $y_j$
when $\rv=\mathbf{0}$ and $\R_j=\mathbf{0}$,
we adopt the convention: 
\begin{equation}
\left \{ \rv=0, \; \R_j=0 \right \}  \Rightarrow x_j = 0, \quad y_j = 0. 
\end{equation}
Finally, we define the hopping matrix elements
\begin{equation}
t^{\alpha\beta,mn}_{\kv}(\rv) = \langle \alpha^{(m)} |  \Hcal_{\kv}^{\rm int}(\rv)| \beta^{(n)}\rangle 
\end{equation}
between orbitals $\alpha,\beta \in \{p_x, p_y, p_z\}$ located at the O atoms labeled by $m, n \in \{1,\ldots,4\}$ in the top and bottom layers, respectively. 

First consider the hopping matrix elements between 
the orbitals on atom O$_3$ of the top layer 
with those of atoms O$_4$ in the bottom layer.
For example, the hopping matrix element between $p_z$ orbitals at the top-layer O$_3$ atoms and bottom-layer O$_4$ atoms is given by:
\begin{equation}
    t^{zz,34}_{\kv} (\rv) = t_0^{zz,34} 
    \sum_{j} e^{ i \kv\cdot\rv_j} 
    e^{-\xi ({\bar r}_j)^2} 
    e^{-\kappa (\delta {\bar z}_j)^2}
    f_{\rm c}({\bar r}_j).
    \label{eq:tofr_zz}
\end{equation}
where $\xi$ and $\kappa$ are 
dimensionless positive constants.
Similarly, the hopping matrix elements 
between $p_x$ and $p_y$ orbitals of the 
top-layer O$_3$ atom and $p$-orbitals 
in the bottom-layer O$_4$ atoms with label $j$ are 
given by:
\begin{equation}
    t^{\g z,34}_{\kv} (\rv) =  t_0^{\g z,34}
    \sum_{j} \g_j  e^{ i \kv\cdot\rv_j} 
    e^{-\xi ({\bar r}_j)^2} 
    e^{-\kappa (\delta {\bar z}_j)^2}
    f_{\rm c}({\bar r}_j), 
    \label{eq:tofr_xz}
\end{equation}
\begin{equation}
    t^{\g\g,34}_{\kv} (\rv) =  t_0^{\g\g,34}
    \sum_{j} \g_j e^{ i \kv\cdot\rv_j} 
     e^{-\xi ({\bar r}_j)^2} 
    e^{-\kappa (\delta {\bar z}_j)^2}
    f_{\rm c}({\bar r}_j),
    \label{eq:tofr_xx} 
\end{equation}
\begin{equation}
    t^{xy,34}_{\kv} (\rv) =  t_0^{xy,34}
    \sum_{j} x_j y_j  e^{ i \kv\cdot\rv_j} 
    e^{-\xi ({\bar r}_j)^2} 
    e^{-\kappa (\delta {\bar z}_j)^2}
    f_{\rm c}({\bar r}_j),
    \label{eq:tofr_xy}
\end{equation}
with $\g = x,y$.
By symmetry, $t^{xz,34}_{0} = t^{yz,34}_{0}$ 
and $t^{xx,34}_{0} = t^{yy,34}_{0}$.
In the above expressions 
for the hopping matrix elements 
we have implicitly made the approximation that only terms 
of $\sigma$-type bonding contribute significantly,
while terms of $\pi$-type bonding can be neglected.

For hopping matrix elements between orbitals at
the top-layer O$_1$ and O$_2$ atoms and those
of the bottom layer O$_4$ and O$_1$, O$_2$ atoms, 
we note that the relevant orbitals 
at the top-layer O$_1$, O$_2$ atoms 
are of $p_z$ character, so the expressions 
are the same as those given  
in Eq.~\eqref{eq:tofr_zz} and Eq.~\eqref{eq:tofr_xz}
but with different values for the parameters involved.
Finally, we ignore the hopping between any orbital 
at the top-layer O$_4$ atom and the bottom-layer 
O$_3$ atom, as those atoms are too far from each other 
to experience any significant interaction. 

\begin{figure*}[t!]
    \centering
    \includegraphics[width=1.0\textwidth]{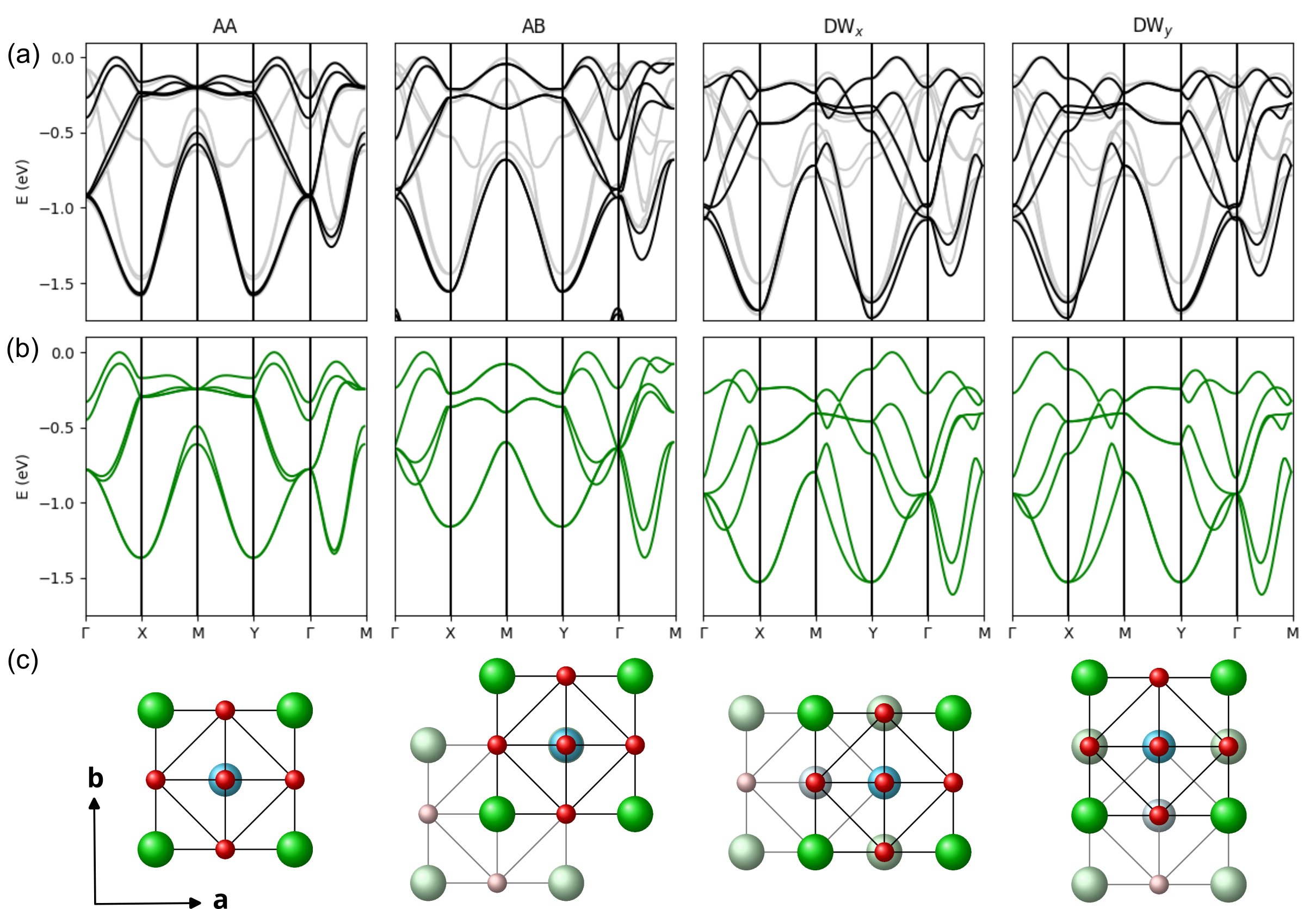}
    \caption{Band structures of STO RP$_1$  bilayers in 4 high symmetry stacking arrangements. {\bf (a)} Valence bands calculated with DFT (grey) and the MLWF model for $\M$-odd valence bands (black) from Eqs.~\eqref{eq:H-tot}-\eqref{eq:Hc} and \eqref{eq:Hint-AA}-\eqref{eq:Hint-DY} with MLWF parameters from Tables~\ref{tab:params} and \ref{tab:inter-mlwf}. {\bf (b)} Simplified 3-band (per layer) model built from the upper 3$\times$3 block of the MLWF model, Eqs.~\eqref{eq:H3}, \eqref{eq:Hr00}-\eqref{eq:Hr01}, using reoptimized parameters as given in Tables~\ref{tab:params-0} and \ref{tab:inter-mlwf-renorm} (green). {\bf (c)} Top view of the RP$_1$ bilayer crystal structure (perpendicular to the plane of the 2D layer) showing the high symmetry stacking configuration for each column.}
    \label{fig:interlayer-model}
\end{figure*}

We focus on the form of $\Hcal_\kv^\mathrm{int}(\rv)$ at the special configurations AA, AB, DW$_x$, and DW$_y$, which are considerably simpler than a general configuration and provide physical insight. 
The interlayer interaction terms for the full 10 band hamiltonian (5 bands from each layer) can be accurately determined using MLWFs. 
The expressions for $\Hcal^{\rm int}_{\kv}(\rv)$ for these configurations are:
\begin{widetext}
\begin{align}
\Hcal^{\rm int}_{\kv} (\rv_{00})&= 
\begin{bmatrix}
u_0 & 0 & -u_1 g_{\kv}(\av_1) & 0 & 0 \\
0 & u_0 & -u_1 g_{\kv}(\av_2) & 0 & 0 \\
u_1 g_{\kv}(\av_1) & u_1 g_{\kv}(\av_2) 
& u_2 + u_3 h_{\kv}(\av_1,\av_2,1,0) 
& 0 & 0 \\
0 & 0 & 0 & 0 & 0 \\
0 & 0 & 0 & 0 & 0
\end{bmatrix},
\label{eq:Hint-AA}
\end{align}
\begin{align}
\Hcal^{\rm int}_{\kv} (\rv_{11})&= 
\begin{bmatrix}
v_1 F_{\kv} 
&
v_2 G_{\kv}
&
-v_3 g_\kv(\frac{\av_1}{2}) f_\kv(\frac{\av_2}{2}) & 0 &
v_5 g_{\kv}(\frac{\av_1}{2}) \\
v_2 G_{\kv} 
&
v_1 F_{\kv} 
&
-v_3 f_\kv(\frac{\av_1}{2}) g_\kv(\frac{\av_2}{2}) &
v_5 g_{\kv}(\frac{\av_2}{2}) & 0 \\
v_3 g_\kv(\frac{\av_1}{2}) f_\kv(\frac{\av_2}{2}) &
v_3 f_\kv(\frac{\av_1}{2}) g_\kv(\frac{\av_2}{2}) &
v_4 F_{\kv} 
&
v_6 f_{\kv}(\frac{\av_2}{2}) & v_6 f_{\kv}(\frac{\av_1}{2}) \\
0 & -v_5 g_{\kv}(\frac{\av_2}{2})  & v_6 f_{\kv}(\frac{\av_2}{2}) & 0 & v_7 \\
-v_5 g_{\kv}(\frac{\av_1}{2}) & 0 & v_6 f_{\kv}(\frac{\av_1}{2}) & v_7 & 0
\end{bmatrix}, 
\label{eq:Hint-AB}
\end{align}

\begin{align}
\Hcal^{\rm int}_{\kv} (\rv_{10}) &= 
\begin{bmatrix}
w_0 f_\kv(\frac{\av_1}{2}) & 0 & -w_2 g_{\kv}(\frac{\av_1}{2}) & -w_8 g_{\kv}(\av_1) & 0 \\
0 & w_1 f_\kv(\frac{\av_1}{2}) & 0 & -w_3 g_\kv(\av_2) & 0 \\
w_2 g_{\kv}(\frac{\av_1}{2}) & 0 
& w_4 f_\kv(\frac{\av_1}{2}) 
& w_5 + w_6 h_{\kv}(\av_1,\av_2,\lambda_D,0) 
&  w_9 F_{\kv} \\
w_8 g_{\kv}(\av_1) & w_3 g_\kv(\av_2) 
& w_5 + w_6 h_{\kv}(\av_1,\av_2,\lambda_D,0) 
& 0 & 0 \\
0 & 0 & w_9 F_{\kv} 
& 0 & 0
\end{bmatrix}, 
\label{eq:Hint-DX}
\end{align}
       
\beq{}
\Hcal^{\rm int}_{\kv} (\rv_{01}) = 
\begin{bmatrix}
w_1 f_\kv(\frac{\av_2}{2}) & 0 & 0 & 0 & -w_3 g_\kv(\av_1) \\
0 & w_0 f_\kv(\frac{\av_2}{2}) & -w_2 g_{\kv}(\frac{\av_2}{2}) & 0 & -w_8 g_{\kv}(\av_2) \\
0 & w_2 g_{\kv}(\frac{\av_2}{2}) 
& w_4 f_\kv(\frac{\av_2}{2}) 
& w_9 F_{\kv} 
& w_5 + w_6 h_{\kv}(\av_2,\av_1,\lambda_D,0) \\
0 & 0 & w_9 F_{\kv} 
& 0 & 0 \\
 w_3 g_\kv(\av_1) & w_8 g_{\kv}(\av_2) 
& w_5 + w_6 h_{\kv}(\av_2,\av_1,\lambda_D,0)
& 0 & 0 
\end{bmatrix},
\label{eq:Hint-DY}
\eeq
\end{widetext}
where we have defined 
\begin{equation}
F_{\kv} = f_{\kv}\left (\frac{\av_1}{2} \right ) f_{\kv}\left (\frac{\av_2}{2}\right), \quad 
G_{\kv} = g_{\kv}\left (\frac{\av_1}{2} \right ) g_{\kv}\left (\frac{\av_2}{2}\right ),
\label{eq:FG-definition}
\end{equation}
with $f_{\kv}(\av)$ and $g_{\kv}(\av)$ defined in Eqs.~(\ref{eq:fk-definition})-(\ref{eq:gk-definition}) and
$h_{\kv}(\av,\av',\lambda,\mu)$ defined in 
Eq.~\eqref{eq:hk-definition}.
The parameter values for CTO, STO, and BTO are listed in Table~\ref{tab:inter-mlwf}, with $\lambda_D = w_7/w_6$. 
The resulting bands for STO are shown in the top row of Fig.~\ref{fig:interlayer-model} where they are compared to the bands obtained from DFT calculations. 
For each stacking arrangement the interlayer interactions result in a VBM consisting of $\mathcal{M}$-odd states away from $\Gamma$, validating our choice to focus on the $\mathcal{M}$-odd states.
Similar plots for CTO and BTO are included as Figs.~S6 and S7.
As an example of the general formalism for the hopping matrix elements, we work out a few special cases in detail in Appendix~\ref{sec:appendix-interlayer-example}. 
\begin{table*}[t!]
  \centering
  \caption{The numerical parameters based on MLWF for the interlayer hamiltonians in Eqs.~\eqref{eq:Hint-AA}-\eqref{eq:Hint-DY}, in units of eV.}
\begin{tabular}{|l|r|r|r|r|r|r|r|r|r|r|r|}
  \hline
   & $u_0$ & $u_1$ & $u_2$ & $u_3$ & $v_1$ & $v_2$ & $v_3$ & $v_4$ & $v_5$ & $v_6$ & $v_7$ \\  \hline
  CTO & 0.011 & 0.005 & 0.069 & 0.006  & $-0.022$ & $-0.057$ & 0.040 & 0.040 & 0.033 & 0.028 & 0.017 \\
STO & 0.007 & 0.003 & 0.058 & 0.005  & $-0.007$ & $-0.027$ & 0.025 & 0.025 & 0.036 & 0.037 & 0.019  \\
BTO & 0.002 & 0.001 & 0.032 & 0.002  & 0.007 & $-0.002$ & 0.006 & 0.004 & 0.035 & 0.048 & 0.023  \\
  \hline\hline
   &  $w_0$ & $w_1$ & $w_2$ & $w_3$ & $w_4$ & $w_5$ & $w_6$ & $w_7$ & $w_8$ & $w_9$ & \\  \hline
  CTO & $-0.042$ & 0.042 & 0.119 & 0.010 & 0.148 & $-0.005$ & 0.017 & 0.014 & 0.008 & 0.008 & \\
STO & $-0.026$  & 0.017 & 0.078 & 0.012 & 0.117 & $-0.007$ & 0.012 & 0.020 & 0.006 & 0.009 & \\
BTO & $-0.012$ & $-0.004$ & 0.031 & 0.011 & 0.073 & $-0.008$ & 0.006 & $0.025$ & 0.005 & 0.010 & \\
\hline
  \end{tabular}
  \label{tab:inter-mlwf}
\end{table*} 

In the spirit of reducing the number of relevant bands as much as possible, we again focus on the three bands that form the top of the valence band manifold. The following simplified interaction hamiltonians contain the dominant effects of the interlayer interactions:
\begin{align}
\Hcal^{\rm int}_{\kv} (\rv_{00}) &= 
\begin{bmatrix}
0 & 0 & 0 \\
0 & 0 & 0  \\
0 & 0 & u_z \\
\end{bmatrix}, \label{eq:Hr00} \\
\Hcal^{\rm int}_{\kv} (\rv_{11}) &= 
\begin{bmatrix}
0 & v_{xy} G_{\kv}
& v_{xz} J_{\kv}(\av_2,\av_1) \\
v_{xy} G_{\kv}
& 0 & v_{yz}J_{\kv}(\av_1,\av_2) \\
v_{xz}J_{\kv}(\av_2,\av_1) & v_{yz}J_{\kv}(\av_1,\av_2) & v_z F_{\kv}
\end{bmatrix}, \\
\Hcal^{\rm int}_{\kv} (\rv_{10}) &= 
\begin{bmatrix}
0 & 0 & -w_{xz} g_{\kv}(\frac{\av_1}{2}) \\
0 & 0 & 0 \\
w_{xz} g_{\kv}(\frac{\av_1}{2}) & 0 & w_z f_\kv(\frac{\av_1}{2}) 
\end{bmatrix}, \\
\Hcal^{\rm int}_{\kv} (\rv_{01}) &= 
\begin{bmatrix}
0 & 0 & 0  \\
0 & 0 & -w_{yz} g_{\kv}(\frac{\av_2}{2})  \\
0 & w_{yz} g_{\kv}(\frac{\av_2}{2}) & w_z f_\kv(\frac{\av_2}{2}) 
\end{bmatrix}, \label{eq:Hr01}
\end{align}
where
\begin{equation}
J_{\kv}(\av,\av') = f_{\kv}\left(\frac{\av}{2}\right) g_{\kv}\left(\frac{\av'}{2}\right).
\label{eq:J-definition}
\end{equation}
The symmetry between $x$- and $y$-directions implies $v_{xz}=v_{yz}$ and $w_{xz}=w_{yz}$. 
The non-zero entries in the above hamiltonians represent the dominant terms that emerge from the MLWF tight-binding hamiltonian. 
From the standpoint of the simplified model with 3 bands per layer, the coefficients should be thought of as renormalized parameters that take into account the couplings with the deeper valence bands derived from the $p_z^{(i)}$ orbitals of both layers. 
These parameters were optimized by hand and are reported in Table~\ref{tab:inter-mlwf-renorm}. 
The corresponding band structures are shown in the bottom row of Fig.~\ref{fig:interlayer-model}.
\begin{table}[ht]
  \caption{The optimal numerical parameters for the interlayer hamiltonians in Eqs.~\eqref{eq:Hr00}-\eqref{eq:Hr01}.
  }
  \begin{tabular}{|l|r|r|r|r|r|r|}
  \hline
   & $u_z$ & $v_{xy}$ & $v_z$ & $v_{xz}$ & $w_{xy}$ & $w_z$ \\  \hline
  CTO & 0.07 & $-0.06$ & 0.04 & $0.020$ & 0.12 & 0.15 \\
STO & 0.06 & $-0.04$ & 0.05 & $0.005$ & 0.09 & 0.14  \\
BTO & 0.03 & $-0.02$ & 0.04 & $0.001$ & 0.03 & 0.07  \\
  \hline
  \end{tabular}
  \label{tab:inter-mlwf-renorm}
\end{table} 

\section{Discussion}

Twisted RP$_1$ oxide layers represent an example of moir\'{e} physics extending beyond the realm of van der Waals structures. 
This is due to their strong interlayer interactions, evidenced by the changes to the single layer band structure of Fig.~\ref{fig:5-bands_0} when in an AB-stacked bilayer (Fig.~\ref{fig:interlayer-model}).
This work represents the first step toward a realistic model incorporating the effects of interlayer coupling on the electronic properties of twisted perovskite bilayers.

The polarization and ferroelectricity of perovskites are interesting and important properties.
In this work we have modeled the interlayer stacking environments of highly symmetric structures, so the Ti atoms remain coplanar with O$_1$ and O$_2$, resulting in no net polarization.
Twisted layers in commensurate supercell models exhibit internal relaxation of the ionic coordinates.
Modeling this atomic relaxation and the resultant electronic structure is the subject of current work.

The present study focused on the valence band, 
which requires detailed modeling of many interacting states that can be experimentally accessed via hole doping, for which we have sought to provide a complete picture. 
A separate description is required for the conduction band that arises primarily from the titanium $d_{xy}$ orbitals and can be experimentally accessed via electron doping.
Construction of a conduction band model is currently underway.

The analysis presented in this paper of the electronic structure of RP$_1$ perovskite bilayers will serve as the foundation for theoretical and numerical models of twisted oxide interfaces.
Of the 12 states in the valence band manifold derived from oxygen $p$-orbitals, we have identified and built an effective model for the three states that play a dominant role in the interlayer interactions between two layers in various stacking arrangements. 
The parametrization of the interlayer coupling in configuration space, namely as a function of the relative shift between two unrotated layers, as presented in Sec.~\ref{sec:interlayer}, provides the necessary ingredients for full-scale tight-binding models of twisted layers that contain all the different local environments. 
This is analogous to what has been done in the case of graphene \cite{fang2016electronic} and TMDs \cite{fang2015ab}.

Continuum models are a complementary method for modeling twisted structures. 
In this approach the band structure for arbitrary twist angles is expanded in terms of moir\'{e} Bloch states that are coupled by the moir\'{e} potential, which can be extracted from the variation of interlayer band energy splittings over configuration space. 
Combining that with the locations of the valence band maxima in the Brillouin zone of the bilayer, we can construct a continuum model as was done for TMDs \cite{angeli2021gamma,angeli2022twistronics}, which will be the subject of future work.

\section*{Acknowledgments}
The computations in this work were run on the FASRC Cannon cluster supported by the FAS Division of Science Research Computing Group at Harvard University. 
We acknowledge financial support from the Simons Foundation Award No.~896626, the US Army Research Office MURI project under Grant No.~W911NF-21-1-0184, with funding from the Army Educational Outreach Program (AEOP).

\appendix 

\section{First-principles calculations}
\label{sec:dft}

DFT band structure calculations were performed using the VASP code~\cite{kresse1996efficient,kresse1996efficiency}, PBE exchange-correlation functional~\cite{perdew1996generalized}, and projector augmented wave (PAW) pseudopotentials with 10, 10, and 6 valence electrons for the A, B, and O atoms, respectively. 
For all three materials studied we employed a plane-wave energy cutoff of 520 eV as well as the zero-damping DFT-D3  van der Waals correction~\cite{grimme2010consistent}.
Forces on each atom were reduced below 0.01 eV/\AA{} during structural relaxations, and we used a $\G$-centered 8$\times$8$\times$1 k-point grid for the final self-consistent calculations. The maximally localized Wannier transformations were performed using the Wannier90 code~\cite{mostofi2008wannier90}, with initial projections onto the oxygen $p$-orbitals.

\section{Fitting the GSFE to smooth functions}
\label{sec:appendix-GSFE}

The GSFE and GSFH act like scalars in 2D and are therefore even functions of $\rv = (x,y)$.
Because they are periodic, we can expand them in a basis of Fourier harmonics:
\beq{}
f(x,y) = c_0 + \sum_{n=1}^N c_n f_{n}^{(1)}(x,y) + d_n f_{n}^{(2)}(x,y) 
\eec

\beq{}
\begin{split}
f_{n}^{(1)}(x,y) &= 
\cos(2\pi n x) + \cos(2\pi n y) \\
f_{n}^{(2)}(x,y) &= 
\cos(2\pi n x) \cos(2\pi n y)
\end{split}
\eec
where the coefficients $c_n$ and $d_n$ are given in Table~\ref{tab:GSFE} for the GSFE and in Table~\ref{tab:GSFH} for the GSFH.
\begin{table}[ht]
\caption{GSFE fitting parameters, in eV.}
\begin{tabular}{|l|r|r|r|r|r|}
  \hline
   & $c_0$ & $c_1$ & $c_2$ & $c_3$ & $c_4$ \\  \hline
  CTO & 1.525 & 0.533 & $-0.0870$ & $-0.00208$ & 0.00297  \\
  STO & 1.097 & 0.389 & $-0.0608$ & $-0.00012$ & 0.00215  \\
  BTO & 0.656 & 0.235 & $-0.0381$ & $ 0.00016$ & 0.00113  \\
  \hline\hline
  & & $d_1$ & $d_2$ & $d_3$ & $d_4$ \\  \hline
  CTO & & $-0.394$ & 0.0610 & 0.00696 & $-0.00246$ \\
  STO &  & $-0.259$ & 0.0339 & 0.00533 & $-0.00135$ \\
  BTO &  & $-0.141$ & 0.0170 & 0.00366 & $-0.00181$ \\
  \hline
  \end{tabular}
  \label{tab:GSFE}
\end{table} 

\begin{table}[ht]
\caption{GSFH fitting parameters, in \AA{}.}
\begin{tabular}{|l|r|r|r|r|r|}
  \hline
      & $c_0$ & $c_1$ & $c_2$ & $c_3$ & $c_4$ \\  \hline
  CTO & 6.780 & 0.653 & 0.0493 & $-0.0160$ & $-0.00445$ \\
  STO & 7.057 & 0.594 & 0.0434 & $-0.0176$ & $ 0.00244$  \\
  BTO & 7.477 & 0.552 & 0.0186 & $-0.0280$ & $-0.01490$  \\
  \hline\hline
    & & $d_1$ & $d_2$ & $d_3$ & $d_4$ \\  \hline
  CTO &  & 0.374 & $-0.0762$ & $-0.00795$ & $ 0.00526$ \\
  STO &  & 0.369 & $-0.0293$ & $-0.04380$ & $-0.00961$ \\
  BTO &  & 0.242 & $-0.0012$ & $-0.01980$ & $ 0.01361$ \\

\hline
  \end{tabular}
  \label{tab:GSFH}
\end{table} 

\section{Interlayer matrix elements}
\label{sec:appendix-interlayer-example}

As an example of the general formalism for the hopping matrix elements, we work out a few special cases in detail. 
For the case of $t^{zz,34}_{\kv}(\rv)$, using Eq.~\eqref{eq:tofr_zz} for the AA configuration we obtain:
\begin{align}
    t^{zz,34}_{\kv}(\rv_{00}) & = t_0^{zz,34} e^{-\kappa} \nonumber \\ 
 &    \left [ 1 + e^{-\xi} \left (
    e^{i \kv\cdot\av_1} + e^{-i \kv\cdot\av_1}
    + e^{i \kv\cdot\av_2} + e^{-i \kv\cdot\av_2}
    \right ) \right ]  \nonumber \\ 
    & = t_0^{zz,34} e^{-\kappa} \left [ 1 + e^{-\xi} \left ( 
    f_{\kv}(\av_1) + f_{\kv}(\av_2) \right ) 
    \right ],
\end{align}
which, with the identification 
\begin{equation}
t_0^{zz,34} e^{-\kappa} \equiv u_2, \quad 
t_0^{zz,34} e^{-\kappa} e^{-\xi} \equiv u_3, 
\end{equation}
is identical to the corresponding matrix element 
of $\Hcal_{\kv}^{\rm int}(\rv_{00})$, 
Eq.~\eqref{eq:Hint-AA}.
For the AB configuration, we obtain:
\begin{align}
    t^{zz,34}_{\kv} (\rv_{11}) & = t_0^{zz,34} 
    e^{-\kappa (\delta {\bar z}_{11})^2 } e^{-\xi/2} \nonumber \\ 
   & \left [ e^{i \kv \cdot \frac{\av_1+\av_2}{2}} + 
    e^{-i \kv \cdot \frac{\av_1+\av_2}{2}} +
    e^{i \kv \cdot \frac{\av_1-\av_2}{2}} + 
    e^{-i \kv \cdot \frac{\av_1-\av_2}{2}} 
    \right ] \nonumber \\
    & = t_0^{zz,34} e^{-\kappa (\delta {\bar z}_{11})^2} 
    e^{-\xi/2} 2 f_{\kv}\left( \frac{\av_1}{2} \right ) 
    f_{\kv}\left(\frac{\av_2}{2}\right)
\end{align}
which, with the identification 
\begin{equation}
2 t_0^{zz,34} e^{-\kappa (\delta {\bar z}_{11})^2} 
    e^{-\xi/2} \equiv v_4 
\end{equation}
is identical to the corresponding matrix element 
of $\Hcal_{\kv}^{\rm int}(\rv_{11})$, 
Eq.~\eqref{eq:Hint-AB}.
For the DW$_x$ configuration, we obtain:
\begin{align}
    t^{zz,34}_{\kv} (\rv_{10}) & = t_0^{zz,34} 
    e^{-\kappa (\delta {\bar z}_{10})^2} e^{-\xi/4} 
    \left [ e^{i \kv \cdot \av_1/2} + 
    e^{-i \kv \cdot \av_1/2} \right ] \nonumber \\
    & = t_0^{zz,34} e^{-\kappa (\delta {\bar z}_{10})^2} 
    e^{-\xi/4} 
    f_{\kv}\left ( \frac{\av_1}{2} \right ),
\end{align}
which, with the identification
\begin{equation}
t_0^{zz,34} e^{-\kappa (\delta {\bar z}_{10})^2} 
    e^{-\xi/4} \equiv w_4 
\end{equation}
is identical to the corresponding matrix element 
of $\Hcal_{\kv}^{\rm int}(\rv_{10})$, 
Eq.~\eqref{eq:Hint-DX}.
Similarly, for the DW$_y$ configuration, we obtain: 
\begin{align}
    t^{zz,34}_{\kv} (\rv_{01}) & = t_0^{zz,34} 
    e^{-\kappa (\delta {\bar z}_{01})^2} e^{-\xi/4} 
    \left [ e^{i \kv \cdot \av_2/2} + 
    e^{-i \kv \cdot \av_2/2} \right ] \nonumber \\
    & = t_0^{zz,34} e^{-\kappa (\delta {\bar z}_{01})^2} 
    e^{-\xi/4} 
    f_{\kv}\left ( \frac{\av_2}{2} \right ),
\end{align}
which, with the identification
\begin{equation}
t_0^{zz,34} e^{-\kappa (\delta {\bar z}_{01})^2} 
    e^{-\xi/4} \equiv w_4 
\end{equation}
is identical to the corresponding matrix element 
of $\Hcal_{\kv}^{\rm int}(\rv_{01})$, 
Eq.~\eqref{eq:Hint-DY}, 
since $\delta {\bar z}_{10} = \delta {\bar z}_{01}$
by symmetry. 

\bibliography{./references.bib}

\end{document}


\title{Supplementary Material:
Stacking-dependent electronic structure of ultrathin perovskite bilayers
}

\author{Daniel T.~Larson}
\affiliation{\HarvardPhysics}
\author{Daniel Bennett}
\affiliation{\HarvardSeas}
\author{Abduhla Ali}
\affiliation{\HarvardSeas}
\affiliation{Cooper Union, New York, NY 10003, USA}
\author{Anderson S.~Chaves}
\affiliation{\HarvardSeas}
\author{Raagya Arora}
\affiliation{\HarvardSeas}
\author{Karin M. Rabe}
\affiliation{\Rutgers}
\author{Efthimios Kaxiras} \email{kaxiras@physics.harvard.edu}
\affiliation{\HarvardPhysics}
\affiliation{\HarvardSeas}

\maketitle


\section{Projected Band Structure}

\begin{figure}[h]
\centering
\includegraphics[width=0.7\columnwidth]{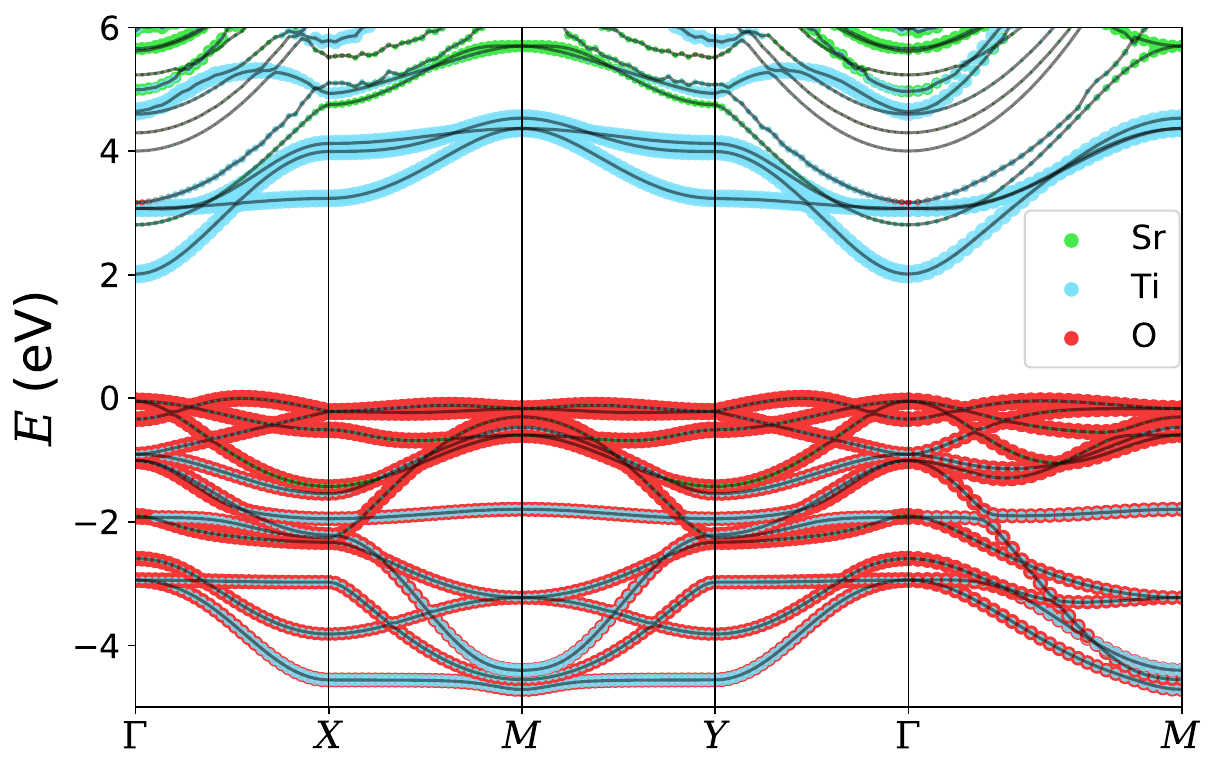}
\caption{Band structure of STO RP$_1$ monolayer showing projections onto atomic species: oxygen (red), titanium (light blue), and strontium (green). As discussed in the main text, the valence bands are primarily derived from oxygen orbitals and the titanium states form the conduction band.
}
\label{fig:STO-proj-bands}
\end{figure}

Fig.~\ref{fig:STO-proj-bands} shows the band structure of the single RP$_1$ layer Sr$_2$TiO$_4$, with colors indicating the projection onto the different atomic species.

\newpage

\section{DFT Wavefunctions}

\begin{figure}[h]
\centering
\includegraphics[width=0.5\columnwidth]{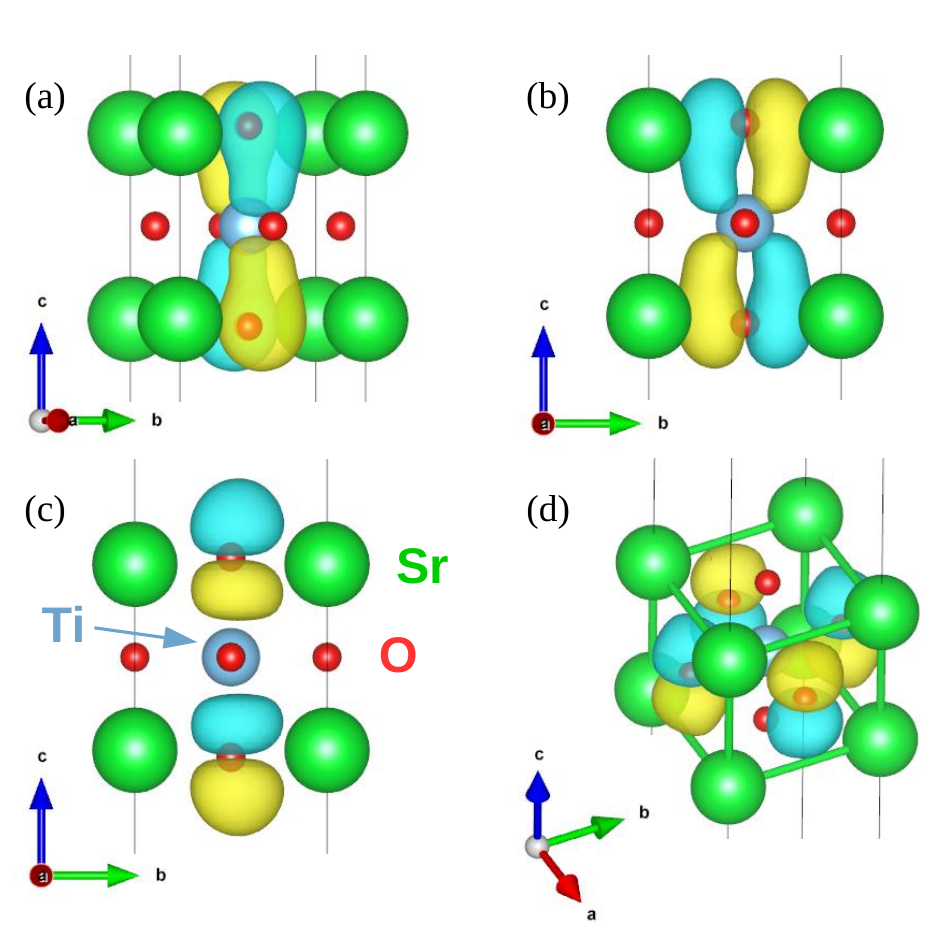}
\caption{Wavefunctions of selected states with well-defined orbital character, 
as obtained from DFT calculations. \textbf{(a)}, \textbf{(b)} The degenerate 
states $\phi_{1,x}$ and $\phi_{1,y}$, with energy just below $-1$ eV at $\Gamma$ in the band structure of Fig.~3(d). (c) The state 
$\psi_{1,z}$ with energy near $-1$ eV at $M$. 
(d) The $\mathcal{M}$-odd state $\chi_{-1,z}$
with energy about $-2$ eV at $\Gamma$.
}
\label{fig:odd-orbs}
\end{figure}

Fig.~\ref{fig:odd-orbs} shows the DFT wavefunctions of selected states with well-defined orbital character for an RP$_1$ layer, Sr$_2$TiO$_4$.

\newpage

\section{GSFE/GSFH of additional materials}

\begin{figure*}[hb!]
\centering
\includegraphics[width=1.0\linewidth]{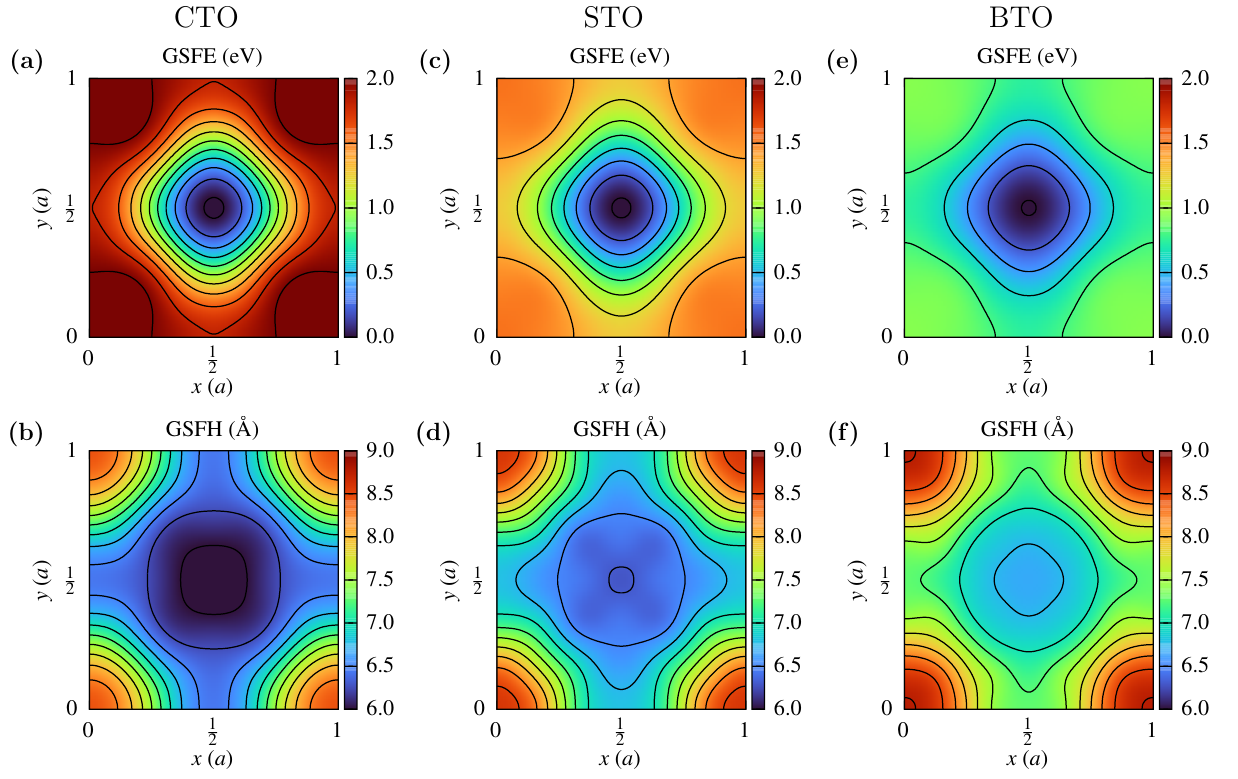}
\caption{
{\bf (a)} Generalized stacking fault energy and 
{\bf (b)} generalized stacking fault height for two STO RP$_1$ monolayers.
Results were obtained from first-principles calculations, including a vdW dispersion correction and allowing for optimization of the distance between the layers, which were then fit to smooth functions (see Appendix B).
}
\label{fig:test-vdw}
\end{figure*}

The GSFE and GSFH for bilayer RP$_1$ structures of CTO, STO and BTO are shown in Fig.~\ref{fig:test-vdw}.

\newpage

\section{Electronic Structure of Additional Materials}

\begin{figure*}[hb]
\centering
\includegraphics[width=.9\textwidth]{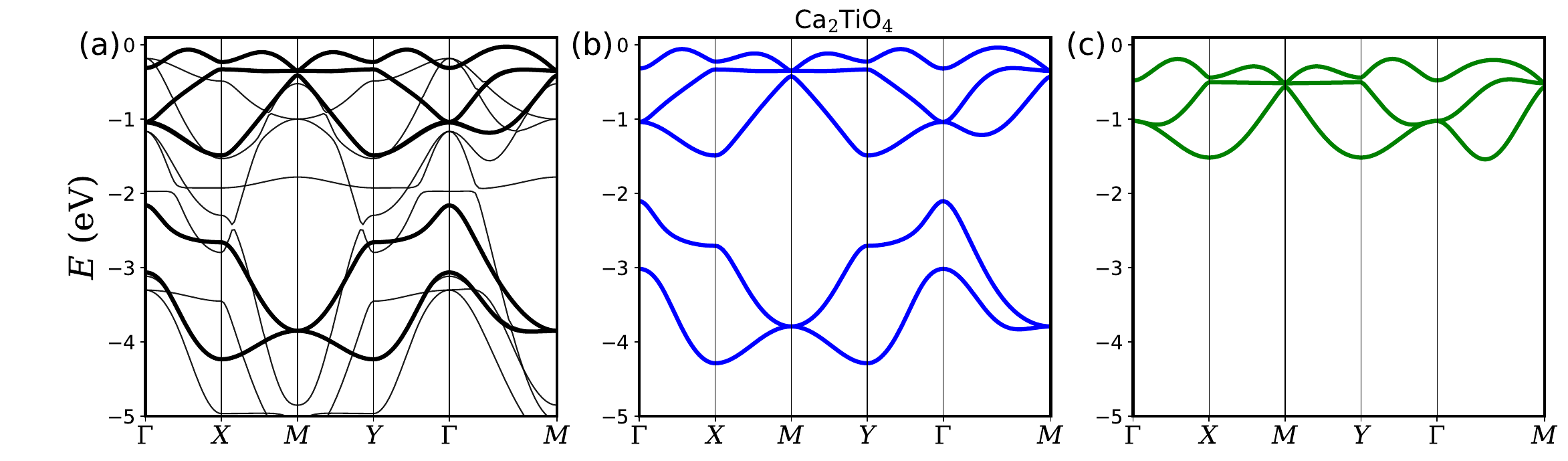}
\caption{CTO RP$_1$ band structure: \textbf{(a)} The 12 valence energy bands calculated using DFT are plotted with thin black lines, with the 5 relevant $\mathcal{M}$-odd bands highlighted in thicker black lines. 
The zero of the energy scale corresponds to the valence band maximum.
\textbf{(b)} The 5 bands in the mirror-odd sector described by Eqs.~(10)-(13) using MLWF parameters from Table~I. 
\textbf{(c)} The energy bands obtained from the simplest model for describing the electronic structure of the CTO RP$_1$ oxide unit, 
namely the hamiltonian in Eq.~(10) with $\mathcal{H}^{(c)}_\mathbf{k}$ set to zero and remaining parameters given in Table~II.
}
\label{fig:5-bands_CTO}
\end{figure*}

\begin{figure*}[hb]
\centering
\includegraphics[width=.9\textwidth]{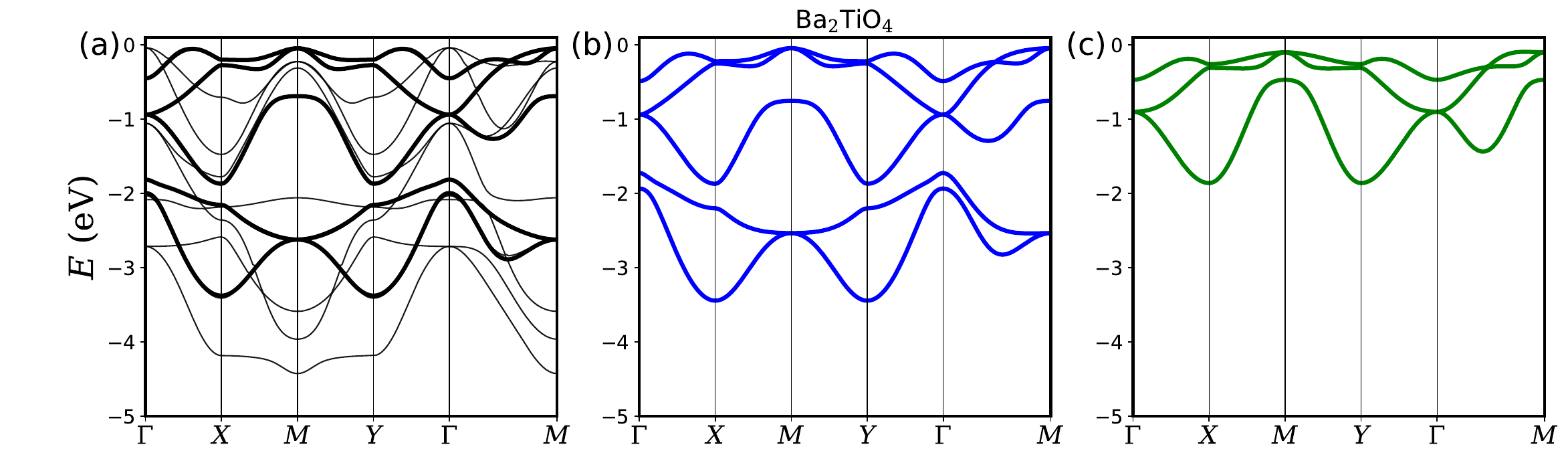}
\caption{BTO RP$_1$ band structure: \textbf{(a)} The 12 valence energy bands calculated using DFT are plotted with thin black lines, with the 5 relevant $\mathcal{M}$-odd bands highlighted in thicker black lines. 
The zero of the energy scale corresponds to the valence band maximum.
\textbf{(b)} The 5 bands in the mirror-odd sector described by Eqs.~(10)-(13) using MLWF parameters from Table~I. 
\textbf{(c)} The energy bands obtained from the simplest model for describing the electronic structure of the BTO RP$_1$ oxide unit, 
namely the hamiltonian in Eq.~(10) with $\mathcal{H}^{(c)}_\mathbf{k}$ set to zero and remaining parameters given in Table~II.
}
\label{fig:5-bands_BTO}
\end{figure*}

\begin{figure*}[ht]
    \centering
    \includegraphics[width=1.0\textwidth]{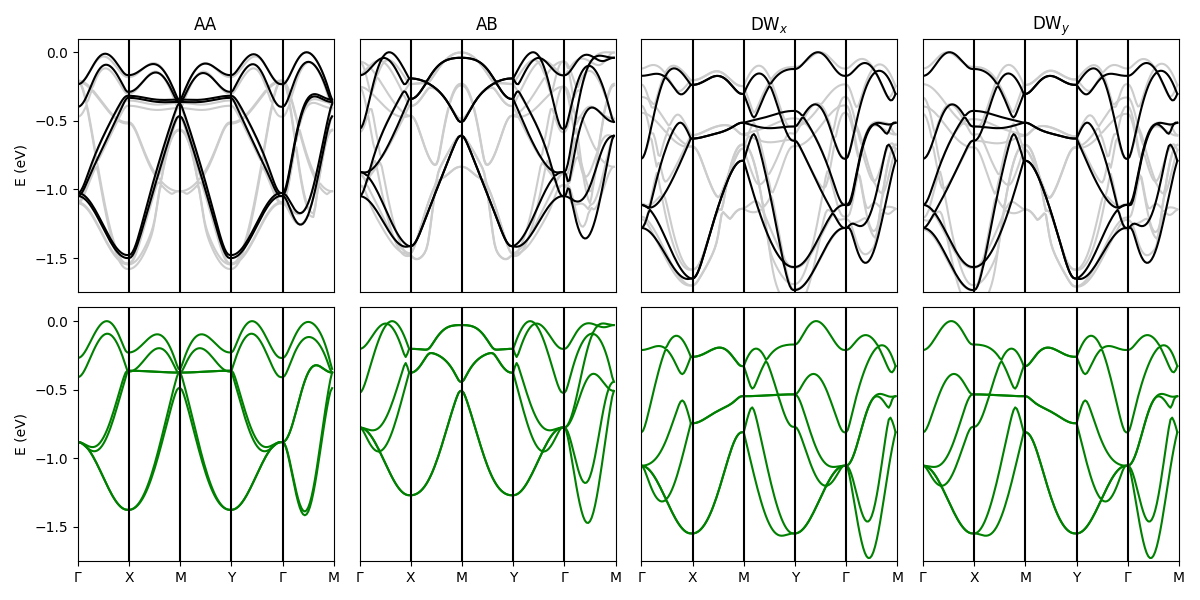}
    \caption{Band structures of CTO RP$_1$  bilayers in 4 high symmetry stacking arrangements. \textbf{(Top)} Valence bands calculated with DFT (grey) and the MLWF model for mirror-odd valence bands (black) from Eqs.~(10)-(13) and (30)-(33) in the main text, with MLWF parameters from Tables~I and III. Note that the avoided crossing in the top 2 bands between X/Y and M for AB stacking is present in all 3 band structures, though it is not visible for the black bands on the scale of this plot. \textbf{(Bottom)} Simplified 3-band (per layer) model built from the upper 3$\times$3 block of the MLWF model, Eqs.~(11), (35)-(38), using reoptimized parameters as given in Tables~II and IV (green).}
    \label{fig:interlayer-CTO}
\end{figure*}

\begin{figure*}[ht]
    \centering
    \includegraphics[width=1.0\textwidth]{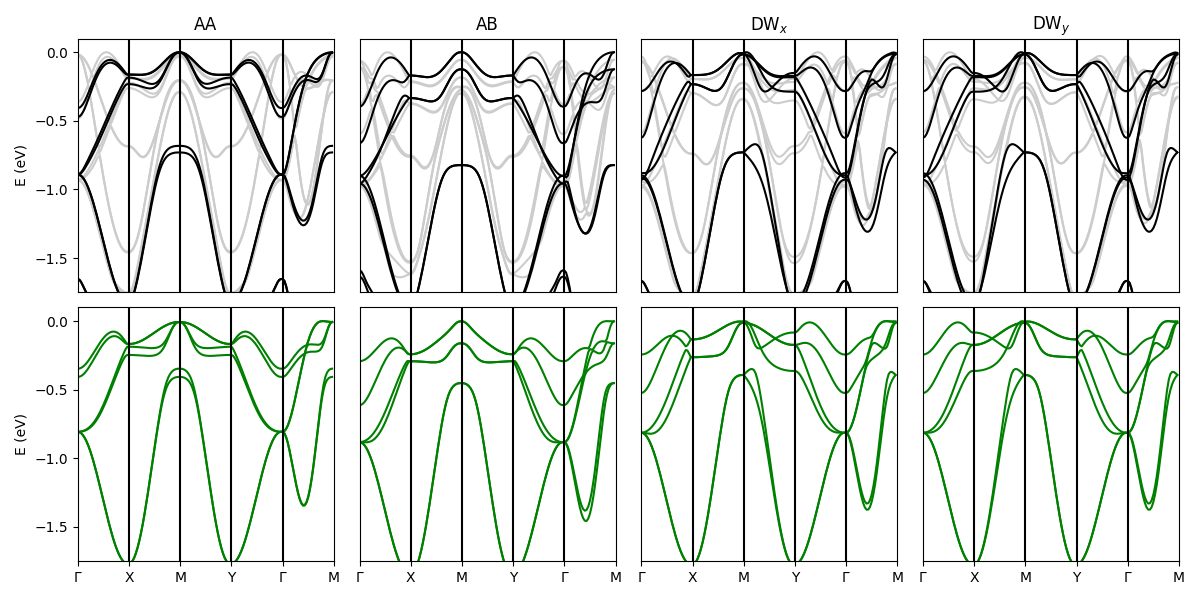}
    \caption{Band structures of BTO RP$_1$  bilayers in 4 high symmetry stacking arrangements. \textbf{(Top)} Valence bands calculated with DFT (grey) and the MLWF model for mirror-odd valence bands (black) from Eqs.~(10)-(13) and (30)-(33) in the main text, with MLWF parameters from Tables~I and III. \textbf{(Bottom)} Simplified 3-band (per layer) model built from the upper 3$\times$3 block of the MLWF model, Eqs.~(11), (35)-(38), using reoptimized parameters as given in Tables~II and IV (green).}
    \label{fig:interlayer-BTO}
    \label{fig:interlayer-model}
\end{figure*}

The main text focused on RP$_1$ mono- and bilayers of STO. Here we provide band structure plots for the RP$_1$ monolayers (Figs.~\ref{fig:5-bands_CTO} and \ref{fig:5-bands_BTO}) and bilayers (Figs.~\ref{fig:interlayer-CTO} and \ref{fig:interlayer-BTO}) of CTO and BTO, respectively.

